\documentclass[a4paper,12pt]{article}
\usepackage{geometry}
\usepackage[T1]{fontenc}
\usepackage[utf8]{inputenc}
\usepackage[english]{babel}
\usepackage{mathtools}
\usepackage{amsmath}
\usepackage{amssymb}
\usepackage{xcolor}
\usepackage{graphicx}
\usepackage{nicefrac}
\usepackage[backend=bibtex,sorting=none]{biblatex}
\usepackage{color, colortbl}
\usepackage{algorithm,algorithmic}
\usepackage{hyperref} %
\usepackage{caption}
\usepackage{placeins}
\usepackage{tikz}
\usepackage{subcaption} 
\usepackage{pgfplots}
\usepackage{datetime}
\usepackage{tikz}
\usepackage{tabularx}

\newcommand{\imagi}{\mathrm{i}} %
\newcommand{\dxdv}{\,\mathrm{d}x \mathrm{d}v } %
\newcommand{\dx}{\,\mathrm{d}x }
\newcommand{\dv}{\,\mathrm{d}v }

\newcommand{\Ampere}{Amp\`{e}re}

\DeclarePairedDelimiter\parentheses{\lparen}{\rparen}
\newcommand{\expbb}[1]{\operatorname{exp} \parentheses*{#1}}

\floatname{algorithm}{Algorithm}

\providecommand{\keywords}[1]
{
  \small	
  \textbf{\textit{Keywords---}} #1
}

\author{
\vspace{.5em}\\
\large{Jakob Ameres$^{1,2}$} \\
\vspace{.5em}\\
$^1$\normalsize{Technische Universit\"at M\"unchen, Zentrum Mathematik}\\
\normalsize{Boltzmannstra\ss{}e 3, 85748 Garching, Deutschland}%
\vspace{.5em}\\
$^2$\normalsize{Max-Planck-Institut f\"ur Plasmaphysik}\\
\normalsize{Boltzmannstra\ss{}e 2, 85748 Garching, Deutschland}%
\vspace{.5em}\\
}
\title{Splitting methods for Fourier spectral discretizations of the strongly magnetized Vlasov--Poisson and the Vlasov--Maxwell system}

\begin{document}
\maketitle

\begin{abstract}
Fourier spectral discretizations belong to the most straightforward methods for solving the unmagnetized Vlasov--Poisson system in low dimensions. In this article, this highly accurate approach is extended two the four-dimensional magnetized Vlasov--Poisson system with new splitting methods suited for strong magnetic fields. Consequently, a comparison to
the asymptotic fluid model is provided at the example of a turbulent Kelvin--Helmholtz instability. 
For the three dimensional electromagnetic Vlasov--Maxwell system different novel charge conserving implementations of a Hamiltonian splitting are discussed and simulation results of the Weibel streaming instability are presented.
\end{abstract}
\keywords{Vlasov--Poisson, Vlasov--Maxwell, strongly magnetized, Spectral methods, Splitting schemes}

\FloatBarrier
\section{Introduction}
\begin{figure}[H]
\centering
 \includegraphics[width=0.6\textwidth]{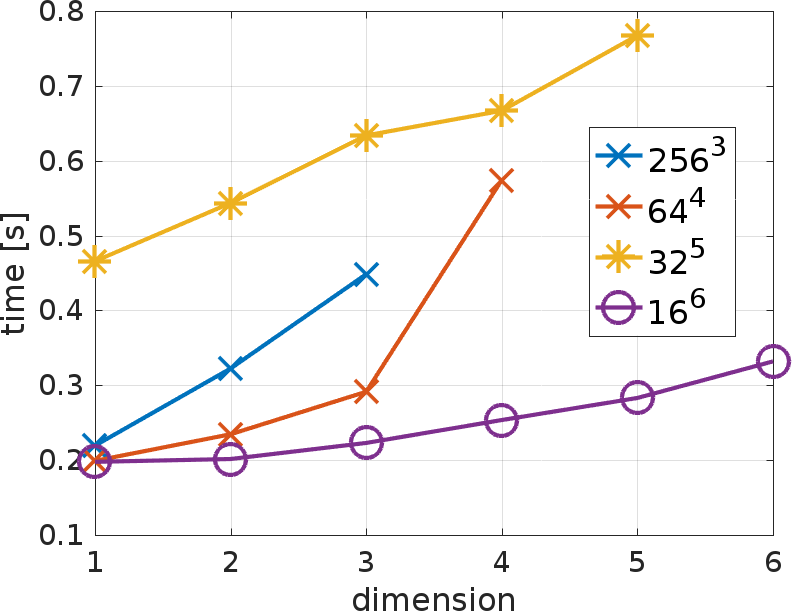}
 \caption{Fourier transforming a multidimensional array along one particular dimension yields a strided access pattern resulting in a slowdown.
          Timings are shown for forth- and back-transform in MATLAB (using FFTW) on a laptop. 
          Although a slowdown is visible, it is not prohibitive for high dimensional spectral methods.}
 \label{spectral:fft_multidim_scaling}
\end{figure}

The Vlasov equation can be discretized following a Lagrangian or Eulerian approach. Lagrangian particle methods such as Particle in Cell
have been dominant for a long time because they share the characteristics with actual physical particles and are easy to implement and parallelize~\cite{birdsall2004plasma}. Although their convergence rate is strongly limited by the Monte-Carlo approach and they based on moving large amounts of data in memory (the particles) it is possible to yield excellent conservation properties~\cite{kraus2016gempic}. 
Semi-Lagrangian methods still use the particles for transporting the distribution function but yield higher convergence rates with an intermediate interpolation step using an Eulerian grid, yet their conservative form remains expensive~\cite{crouseilles2010conservative,crouseilles2014new}. There exists a variety of Eulerian Vlasov--Poisson solvers~\cite{filbet2001conservative} where lately geometric methods gained popularity~\cite{kraus2013variational}. One of the simplest Eulerian solvers are pseudo-spectral solvers.
They, of course, suffer from the curse of dimensionality but not on the computational level here, since the FFTW library is well optimized, see fig.~\ref{spectral:fft_multidim_scaling}.
Constant coefficient advection in a periodic domain can be solved exactly in Fourier space. In all cases treated here, there is a Hamiltonian splitting available yielding constant advection possible. Fourier spectral solvers for the Vlasov equation,
that employ also a Fourier transform in velocity space date back to~\cite{joyce1971numerical,izrar1989integration}. 
Such Fourier-Fourier solver were further developed for higher dimensions \cite{eliasson2001outflow,eliasson2003numerical} and also extended to the Vlasov--Maxwell equation
\cite{eliasson2003numerical,eliasson2010numerical},\cite{crouseilles2015hamiltonian}. It is also possible to use the Fourier basis as the interpolator underlying a Semi-Lagrangian scheme~\cite{FATONE2019349}.
For Vlasov--Poisson it has been shown that Fourier filtering can be used to suppress the recurrence phenomenon \cite{einkemmer2014strategy} or 
filter filamentations \cite{klimas1994splitting}. For Vlasov--Poisson the Hamiltonian splitting has also been known \cite{watanabe2005vlasov},
but for Maxwell, none of these splitting methods is of geometric origin.\\
It should be mentioned that for the velocity space discretization also Chebyshev and Hermite polynomials have been used~\cite{joyce1971numerical,vencels2016spectralplasmasolver}.
There the discretization by low degree Hermite polynomials provides an elegant way to approximate a fluid model on the numerical level. \\
A priori structure should be conserved for long terms and e.g. energy conservation is just a consequence but not the goal itself.
Fourier spectral methods do not conserve positivity of the distribution function. In this context, we neglect the question on positivity conserving schemes
although for other forms of discretizations there have been improvements in that direction~\cite{filbet2001conservative,filbet2003comparison,arber2002critical}.\\
We begin by recalling the mechanisms for the Fourier spectral discretization of the two-dimensional Vlasov--Poisson and Vlasov--Amp\`{e}re systems.
In the second part, the four-dimensional magnetized Vlasov--Poisson system is obtained by the introduction of an external homogeneous magnetic field. There the Fourier spectral counterparts of known exponential splitting methods~\cite{knapp2015splitting} are presented and their performance is investigated under a stronger magnetic field with the use of the Kelvin Helmholtz instability.
In the third part, we turn to electromagnetic physics by the means of the three dimensional Vlasov--Maxwell system, where the methods based on a Hamiltonian splitting are discussed at various test-cases following~\cite{kraus2016gempic}. The implementation in MATLAB used for the numerical examples can be found in a repository~\cite{FourierSpectralVlasovGITLAB}.

\FloatBarrier
\subsection{Vlasov--Poisson (1d1v)}
We consider the one dimensional Vlasov equation~\eqref{spectral:vlasov1d}
\begin{equation}
\label{spectral:vlasov1d}
\partial_t f(x,v,t) + v \partial_x f(x,v,t) + \frac{q}{m} \left(E(x,t) + E_{ext}(x,t)\right) \partial_v f(x,v,t) =0
\end{equation}
and the Poisson equation
\begin{equation}
\label{spectral:poisson1d}
 - \partial_{xx} \Phi(x,t) = 1 + q \int_{-\infty}^\infty f(x,v,t)\dv, \quad  E(x,t) = -\partial_x \Phi(x,t)
\end{equation}
Here we Fourier transform in velocity and spatial space where
$\hat{f}$ denotes a transformation. For notational simplicity the transformed dimension is indicated by $k_x$ or $k_{v}$ in the argument.
The spatial, velocity and fully Fourier transformed densities are defined as
\begin{align}
 \hat{f}(k_x,v,t) &= \frac{1}{L} \int_0^L f(x,v,t) \exp{-\imagi x k_x} \dx, \\
 \hat{f}(x,k_v,t) &=  \frac{1}{v_{\max}-v_{\min}} \int_{v_{\min}}^{v_{\max}} f(x,v,t) \expbb{-\imagi (v-v_{\min}) k_v} \dv, \\
 \hat{f}(k_x,k_v,t) &= \frac{1}{L} \frac{1}{v_{\max}-v_{\min}}  \int_{v_{\min}}^{v_{\max}}\int_0^L f(x,v,t) \expbb{-\imagi (x k_x + (v-v_{\min}) k_v)} \dxdv,
\end{align}
where the wave vectors are $k_x=n \frac{2\pi}{L}$ and $k_v = \frac{2\pi}{v_{\max}-v_{\min}}$ for $n\in \mathrm{Z}$. 
Note that one can easily by a Fourier forth and back-transform switch between those three representations on a discrete level.
We split the integration in three parts in $\tau [0,t]$, where the Vlasov steps can be integrated exactly in Fourier space.
\begin{enumerate}
 \item Advection in $x$ 
 \begin{equation}
 \partial_t f(x,v,t) + v \partial_x f(x,v,t) =0
 \end{equation}
 \item Advection in $v$ and Poisson solve
 \begin{equation}
   \partial_t f(x,v,t) + \frac{q}{m} \left(E(x,0) + E_{ext}(x,0)\right) \partial_v f(x,v,t)=0
 \end{equation}
 Here we solve the Poisson equation with constant background (for $q=-1$), but other fields are also possible.
\begin{equation}
 \partial_x E(x,t)= 1 + q \int f(x,v,t) \dv 
\end{equation}
\end{enumerate}
For the splitting we consider the time $[0,t]$ to be one time step.
\begin{enumerate}
\item Advection in $x$ in spatially transformed space\\
Considering $v$ to be a fixed parameter the constant coefficient advection yields an ODE for each Fourier coefficient
 \begin{equation}
 \label{spectral:vp:ccaODE}
 \partial_t \hat{f}(k_x,v,t) = - v  \imagi k_x \hat{f}(k_x,v,t),
 \end{equation}
   which can be solved exactly over this splitting step:
 \begin{equation}
 \label{spectral:vp:ccaODE2}
 \begin{split}
   \hat{f}(k_x,v,t) %
                    &= \hat{f}(k_x,v,0) \expbb{ - v  \imagi k_x t }.
 \end{split}
 \end{equation}
 \item  Advection in $v$ in velocity transformed space
 \begin{equation}
 \begin{split}
 \partial_t \hat{f}(x,k_v,t) &=    - \frac{q}{m} \left(E(x,0) + E_{ext}(x,0)\right)   \imagi k_v    \hat{f}(k,k_v,t)  \\
 \hat{f}(x,k_v,t) &=   \hat{f}(x,k_v,0) \expbb{ - \frac{q}{m} \left(E(x,0) + E_{ext}(x,0)\right)   \imagi k_v t }.
 \end{split}
 \end{equation}
 Note that in this step the advection in $v$ cancels out under the velocity integral.
 \begin{equation}
 \begin{split}
  \int_{\mathbb{R}} f(x,v,t) \,\mathrm{d}v &= 
  \int_{\mathbb{R}} f(x,v+ t \frac{q}{m} \left[ E(x,0) + E_{ext}(x,0)\right] ,0) \,\mathrm{d}v   
  =\int_{\mathbb{R}} f(x,v,0) \,\mathrm{d}v
  \end{split}
 \end{equation}
 Therefore, the electric field can be obtained in the spatially transformed space before or at the end of the split step.
 \begin{equation}
 \hat{E}(k_x,0) = q \frac{1}{\imagi k_x} \int \hat{f}(k_x,v,0) \dv , \text{ for } k_x \neq 0
 \end{equation}
\end{enumerate}
The Lie steps can be composed by symmetric composition, see \cite{watanabe2005vlasov}. The symplectic Runge Kutta scheme from Forest and Ruth \cite{forest1989fourth}
also works as it is just shifted by a half step and, therefore, adjoint symplectic for the Eulerian discretization.

\subsection{Vlasov--Amp\`{e}re (1d1v)}
For the Vlasov--Amp\`{e}re formulation the Poisson equation needs to be solved only once at $t=0$, such that the electric field evolves in time by the Amp\`{e}re equation
\begin{equation}
 \partial_t E(x,t) = - j(x,t) = - \sum_s \frac{q }{m } \int v f_s(x,v) \,\mathrm{d}v.
\end{equation}
This leaves us with the following splitting:
\begin{align}
\label{spectral:vlasovampere:splitA}
 &\begin{cases}
 \partial_t f_s(x,v,t) + v \partial_x f_s(x,v,t) =0  \\
 \partial_t E(x,t)=  - \sum_s \frac{q }{m } \int v f_s(x,v,t) \mathrm{d}v
 \end{cases}\\
 &\begin{cases}
 \partial_t f_s(x,v,t) + \frac{q}{m} \left(E(x,t) + E_{ext}(x,t)\right) \partial_v f_s(x,v,t)=0
 \end{cases} 
  \label{spectral:vlasovampere:splitB}
\end{align}
The second split step \eqref{spectral:vlasovampere:splitB} is now missing the Poisson equation but can be solved as before, whereas the second one
\eqref{spectral:vlasovampere:splitB} incorporates now the Amp\`{e}re equation.
It can be integrated exactly, since the solution to the constant coefficient advection is known to be $f_s(x,v,t) = f_s(x - t v,0)$ which can be inserted into the Amp\`{e}re 
equation reading 
 \begin{equation}
 \label{spectral:vlasovampere:ampere1}
 \begin{split}
  E(x,t)&= E(x,0)   - \sum_s \frac{q }{m } \int_0^t   \int v f_s(x,v,t) \,\mathrm{d}v \mathrm{d}\tau\\
  &=E(x,0)  - \sum_s \frac{q }{m } \int_0^t   \int v f_s(x- \tau v,v,0) \,\mathrm{d}v \mathrm{d}\tau.
 \end{split}
 \end{equation}
In spatially Fourier transformed space eqn.~\eqref{spectral:vlasovampere:ampere1} can be solved by inserting the solution of the constant coefficient advection 
given in \eqref{spectral:vp:ccaODE2} as follows:
\begin{equation}
\label{spectral:vlasovampere:ampere1:fourier}
\begin{split}
  \hat{E}(k_x,t)&= \hat{E}(k_x,0)   - \sum_s \frac{q }{m } \int_0^t   \int v \hat{f}_s(k_x,v,t) \,\mathrm{d}v \mathrm{d}\tau\\
                   &= \hat{E}(k_x,0)   - \sum_s \frac{q }{m }  \int_0^t \int v \hat{f}_s(k_x,v,0) \expbb{ - v  \imagi k_x \tau }    \,\mathrm{d}v \mathrm{d}\tau\\ 
                   & = \hat{E}(k_x,0)   - \sum_s \frac{q }{m }                   
                   \begin{cases}
                            t \int v \hat{f}_s(k_x,v,0) \,\mathrm{d}v                         &\text{for } k_x =0,\\
                              \frac{1}{-   \imagi k_x}  \int  \left[ \expbb{ - v  \imagi k_x t } - 1 \right]  \hat{f}_s(k_x,v,0)  \, \mathrm{d}v  & \text{otherwise}.
                   \end{cases}
\end{split}
\end{equation}
\FloatBarrier
\section{Magnetized Vlasov--Poisson (2d2v)}
The magnetized Vlasov equation reads
\begin{equation}
 \partial_t f_s(x,v,t) + v \cdot \nabla_x f_s(x,v,t) + \frac{q }{m }\left[E(x,t) + v \times B(x,t)  \right] \cdot \nabla_v f(x,v,t)  =0 
\end{equation}
which, reduced to two dimensions for $x=(x_1,x_2),v=(v_1,v_2)$ and the magnetic field $B(x,t) = (0,0,B_3(x,t) )$, reads
 \begin{multline}
  \partial_t f_s(x,v,t) +
  v \cdot \nabla_x f_s(x,v,t)
  + \\
  + \frac{q }{m } \left[ \left( E_1(x,t) +   v_2  B_3(x,t)  \right) \partial_{v_1} f(x,v,t) +
   \left( E_2(x,t)  -v_1  B_3(x,t) \right) \partial_{v_2} f(x,v,t)\right] =0.
 \end{multline}
The canonical Hamiltonian splitting for the magnetized Vlasov--Poisson system reads
 \begin{align}
 \mathcal{H}_{E}
  \begin{cases}
   \partial_t f_s(x,v,t)  + \frac{q }{m }  E(x,t) \cdot \nabla_v f_s(x,v,t)=0\\
   E(x,t) = - \nabla \Phi(x,t),~ -\Delta \Phi(x,t) = \sum_s q_s \int f_s(x,v,t) \,\mathrm{d} v, \\
  \end{cases}  \\
  \label{spectral:VP2d2v:Hp1split}
  \mathcal{H}_{p_1}
  \begin{cases}
   \partial_t f_s(x,v,t)   +   v_1 \partial_{x_1} f_s(x,v,t) -  \frac{q }{m } v_1  B_3(x,t) \partial_{v_2} f(x,v,t)   =0,
  \end{cases}\\
  \label{spectral:VP2d2v:Hp2split}
  \mathcal{H}_{p_2}
  \begin{cases}
   \partial_t f_s(x,v,t)   +   v_2 \partial_{x_2} f_s(x,v,t) +  \frac{q }{m } v_2  B_3(x,t) \partial_{v_1} f(x,v,t)   =0,
  \end{cases}
 \end{align}
but has the disadvantage that spatial Fourier transform in $\mathcal{H}_{p_1}$ and $\mathcal{H}_{p_1}$ requires a convolution between $B_3$ and $f$.
We avoid this by separating the advection in each velocity component.
\begin{align}
\label{spectral:VP2d2v:split2:1}
&  \begin{cases}
   \partial_t f_s(x,v,t)  + \frac{q }{m } \left[  E_1(x,t) +  v_2  B_3(x,t) \right] \partial_{v_1} f_s(x,v,t)=0\\
    E_1(x,t) = - \partial_{x_1} \Phi(x,t),~ -\Delta \Phi(x,t) = \sum_s q_s \int f_s(x,v,t) \,\mathrm{d} v, 
  \end{cases}  \\
  \label{spectral:VP2d2v:split2:2}
 & \begin{cases}
   \partial_t f_s(x,v,t)  + \frac{q }{m } \left[  E_2(x,t) -  v_1  B_3(x,t) \right] \partial_{v_2} f_s(x,v,t)=0\\
    E_2(x,t) = - \partial_{x_1} \Phi(x,t),~ -\Delta \Phi(x,t) = \sum_s q_s \int f_s(x,v,t) \,\mathrm{d} v, 
  \end{cases}  \\
  \label{spectral:VP2d2v:split2:3}
&\begin{cases}
   \partial_t f_s(x,v,t) +   v \cdot \nabla_x f_s(x,v,t)=0
  \end{cases} 
 \end{align}
The Poisson equation in \eqref{spectral:VP2d2v:split2:1} and \eqref{spectral:VP2d2v:split2:2} has precisely the same solution for both split steps since the charge density actually stays constant over
the advection and therefore, needs to be only solved once. If we take a look at the characteristics corresponding to \eqref{spectral:VP2d2v:split2:1}-\eqref{spectral:VP2d2v:split2:3}, 
\begin{align}
&  \begin{cases}
\dot{V_1}(t) &=  \frac{q }{m } \left[ E_1(X(t),t) + V_2(t) B_3(X(t),t) \right]\\
\dot{V_2}(t)&=0\\
\dot{X}(t)&=0\\
  \end{cases}  \\
 & \begin{cases}
\dot{V_1}(t) &= 0\\
\dot{V_2}(t)&= \frac{q }{m }\left[ E_2(X(t),t) - V_1(t) B_3(X(t),t) \right] \\
\dot{X}(t)&=0\\
  \end{cases}  
&\begin{cases}
  \dot{V_1}(t) &= 0\\
\dot{V_2}(t)&= 0 \\
\dot{X}(t)&= V(t) \\
  \end{cases} 
 \end{align}
we realize that the circular gyromotion for a strong magnetic field is not described very well, since it is split along each dimension.
So we desire a spectral counterpart to more robust methods for strong magnetic fields like the exponential Boris algorithm~\cite{hongqin2013,knapp2015splitting}.
\subsection{Exponential splitting}
\begin{figure}
  \begin{tabular}{c c c c }
       \multicolumn{4}{c}{Strang splitting $(120^\circ \leq \theta \leq 180^\circ)$}\\
       \includegraphics[width=0.22\textwidth]{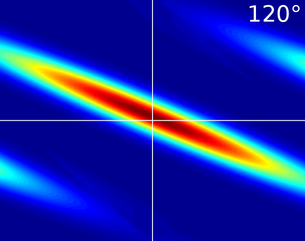} &
       \includegraphics[width=0.22\textwidth]{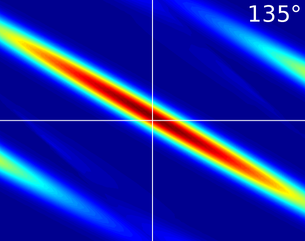} &
       \includegraphics[width=0.22\textwidth]{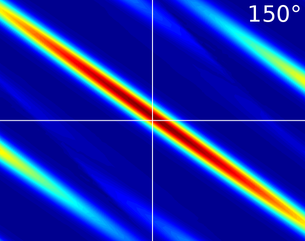} &
       \includegraphics[width=0.22\textwidth]{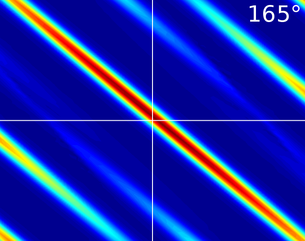} \\
       \multicolumn{4}{c}{shearing $(120^\circ \leq \theta \leq 180^\circ)$}\\
       \includegraphics[width=0.22\textwidth]{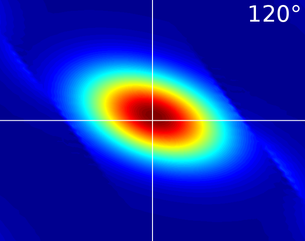} &
       \includegraphics[width=0.22\textwidth]{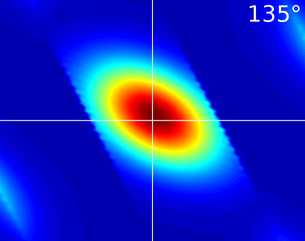} &
       \includegraphics[width=0.22\textwidth]{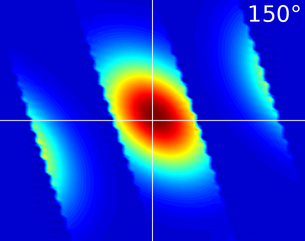} &
       \includegraphics[width=0.22\textwidth]{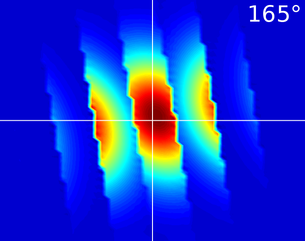} \\
       \hline \\
       \multicolumn{4}{c}{Strang splitting $( -45^\circ \leq \theta \leq 45^\circ)$ and reordering for multiples of $90^\circ$}\\
       \includegraphics[width=0.22\textwidth]{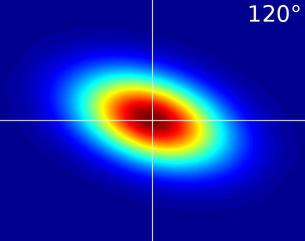} &
       \includegraphics[width=0.22\textwidth]{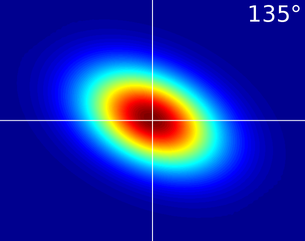} &
       \includegraphics[width=0.22\textwidth]{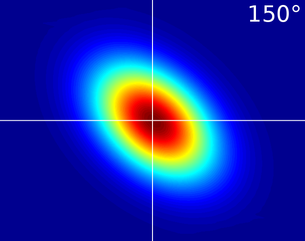} &
       \includegraphics[width=0.22\textwidth]{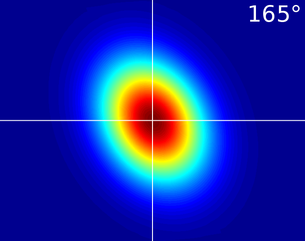} \\
       \multicolumn{4}{c}{shearing $( -45^\circ \leq \theta \leq 45^\circ)$ and reordering for multiples of $90^\circ$}\\
       \includegraphics[width=0.22\textwidth]{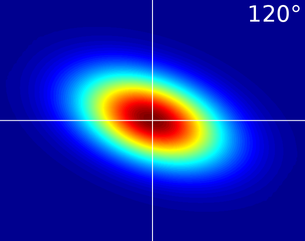} &
       \includegraphics[width=0.22\textwidth]{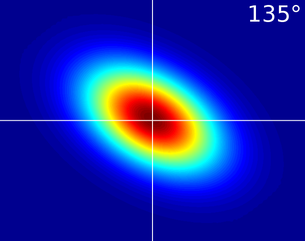} &
       \includegraphics[width=0.22\textwidth]{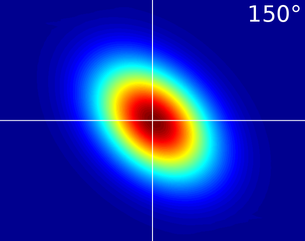} &
       \includegraphics[width=0.22\textwidth]{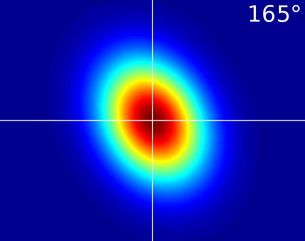} \\
  \end{tabular}
  \caption{Rotating an asymmetric two-dimensional Gaussian (Maxwellian) by Strang splitting and shearing, and with reordering respectively.
           Shearing in Fourier space for more than $90^\circ$ leads to heavy distortions (second row).
           Array rotations by multiples of $90^\circ$ can be implemented exactly by reordering such that shearing in
           Fourier space is only necessary for $-45^\circ \leq \theta \leq 45^\circ$ which leads to much better results (third and fourth row).}
   \label{fig:fourier_rot_maxwellian}
\end{figure}

The characteristics of the splitting underlying the exponential Boris algorithm reads
\begin{align}
&\begin{cases}
  \dot{V}(t) &= 0\\
\dot{X}(t)&= V(t) \\
  \end{cases}\\
&\begin{cases}
  \dot{V}(t) &= \frac{q }{m } E(X(t),t)\\
\dot{X}(t)&= 0 \\
  \end{cases} \\
  \label{eqn:vlasov:char:rot2}
   & \begin{cases}
\dot{V_1}(t) &= \frac{q }{m }  V_2(t) B_3(X(t),t)\\
\dot{V_2}(t)&= - \frac{q }{m } V_1(t) B_3(X(t),t)  \\
\dot{X}(t)&=0\\
  \end{cases} 
 \end{align}
which leads us to the distribution counterpart
\begin{align}
 &\begin{cases}
   \partial_t f_s(x,v,t)  + \frac{q }{m }  E(x,t) \cdot \nabla_v f_s(x,v,t)=0,\\
  \end{cases}  \\
  &\begin{cases}
   \partial_t f_s(x,v,t)   +   v \nabla_x f_s(x,v,t) =0,
  \end{cases}\\
  \label{eqn:vlasov:rot2}
  &\begin{cases} 
   \partial_t f_s(x,v,t)   + \frac{q }{m } \left[ v_2  B_3(x,t)  \partial_{v_1} f(x,v,t) -v_1  B_3(x,t)  \partial_{v_2} f(x,v,t)\right] =0.
  \end{cases}
\end{align}
The exponential Boris scheme, along with many other integrators, use the fact that eqn.~\eqref{eqn:vlasov:char:rot2} can be solved exactly.
With the two-dimensional rotation matrix 
\begin{equation}
 R(\theta)=
 \begin{pmatrix}
  \cos(\theta) & -\sin(\theta) \\
  \sin(\theta) & \cos(\theta)
 \end{pmatrix},
\end{equation}
the solution to eqn.~\eqref{eqn:vlasov:char:rot2} reads
\begin{equation}
  V(t) =  R(-\theta)   V(0)    \text{ for }  \theta= t \frac{q }{m } B_3( X(0),0 ).
\end{equation}
In eqn.~\eqref{eqn:vlasov:rot2} the spatial position $x$ is only a parameter such that the solution to \eqref{eqn:vlasov:rot2} by the methods of characteristics
reads
\begin{equation}
\label{eqn:vlasov:rot2:sol}
f_s(x,v,t)= f_s\left(x, R( - \theta )\cdot v, 0 \right), \quad \theta= t \frac{q }{m } B_3(x,0).
\end{equation}
This corresponds to a rotation in the velocity plane for each position.
In \cite{paeth1986fast} the two-dimensional rotation matrix $R(\theta)$ is decomposed into three shears:
\begin{equation}
 R(\theta)=
 \begin{pmatrix}
  \cos(\theta) & -\sin(\theta) \\
  \sin(\theta) & \cos(\theta)
 \end{pmatrix}
=
 \begin{pmatrix}
  1  & - \tan( \nicefrac{\theta}{2}) \\
  0 & 1 \\
 \end{pmatrix}
 \begin{pmatrix}
  1  & 0 \\
  \sin(\theta) & 1 \\
 \end{pmatrix}
 \begin{pmatrix}
  1  & - \tan( \nicefrac{\theta}{2}) \\
  0 & 1 \\
 \end{pmatrix} 
\end{equation}
Note that the two shears  
\begin{equation}
 S_1 (\alpha) = \begin{pmatrix}
  1 & \alpha\\
  0 & 1
 \end{pmatrix}
 \text{ and }
S_2 (\beta) = \begin{pmatrix}
  1 & 0 \\
  \beta  & 1
 \end{pmatrix}
\end{equation}
Both shears, $S_1$ and $S_2$ correspond merely to a single dimensional advection and can be calculated in Fourier space using one dimensional transforms:
\begin{align}
\mathcal{F}_{v_1}\left[ f(x, S_1(\alpha)\cdot v ) \right]&=    \hat{f}(x, k_{v,1} ,v_2) \expbb{\imagi \alpha v_2  k_{v,1} }, \\
\mathcal{F}_{v_2}\left[ f(x, S_2(\beta) \cdot v ) \right]&=    \hat{f}(x, v_1 ,k_{v,2}) \expbb{\imagi \beta v_1  k_{v,2} }.
\end{align}
This is a commonly known method for image rotation by the discrete Fourier transform~\cite{larkin1997fast}. Contrary to splitting 
this rotation into two sub steps as in \eqref{spectral:VP2d2v:split2:1}-\eqref{spectral:VP2d2v:split2:2} the rotation by shearing
is independent of the relation between time step and magnitude of the potentially strong magnetic field. Using the Taylor expansion for small $\theta$,
by approximating $\sin(x) \approx x $ and $\tan(x) \approx x$ we obtain the shears 
\begin{equation}
 R(-\theta)=
 \begin{pmatrix}
  \cos(\theta) & \sin(\theta) \\
  -\sin(\theta) & \cos(\theta)
 \end{pmatrix}
\approx
 \begin{pmatrix}
  1  &   \nicefrac{\theta}{2} \\
  0 & 1 \\
 \end{pmatrix}
 \begin{pmatrix}
  1  & 0 \\
  -\theta & 1 \\
 \end{pmatrix}
 \begin{pmatrix}
  1  &   \nicefrac{\theta}{2} \\
  0 & 1 \\
 \end{pmatrix}, 
\end{equation}
which correspond exactly to the second order Strang splitting of eqn.~\eqref{eqn:vlasov:rot2}, where the Lie steps read
\begin{equation}
   \Big\{   \partial_t f_s  + \frac{q }{m }  v_2  B_3  \partial_{v_1} f= 0\,   
   \text{ and }
\,\Big\{   \partial_t f_s  - \frac{q }{m }  v_1  B_3  \partial_{v_2} f= 0.
\end{equation}
This also explains why there is no \textit{visible} difference between the Strang splitting and the exact shears in the third and fourth row in fig.~\ref{fig:fourier_rot_maxwellian}.
The rotation of an image multiples of $\nicefrac{\pi}{2}$ can be implemented exactly by permutation involving transposing and flipping arrays, hence
we can restrict the rotation in Fourier space on $\theta \in [-\frac{\pi}{4}, \frac{\pi}{4}]$ as recommended in~\cite{larkin1997fast} and also suggested by fig.~\ref{fig:fourier_rot_maxwellian}.
Another option is to use cubic B-spline interpolation for image rotation which is e.g. provided by \textit{imrotate} in MATLAB. This corresponds to a backward Semi-Lagrangian discretization.\\
In the special case of a homogeneous magnetic field we can consider the splitting underlying Scovel's method:
\begin{align}
  \label{spectral:VP2d2v:split4:3}
&\begin{cases}
  \dot{V}(t) &= \frac{q }{m } E(X(t),t)\\
\dot{X}(t)&= 0 \\
  \end{cases} \\
   & \begin{cases}
\dot{V_1}(t) &= \frac{q }{m }  V_2(t) B_3\\
\dot{V_2}(t)&= - \frac{q }{m } V_1(t) B_3  \\
\dot{X}(t)&=V(t)\\
  \end{cases} 
 \end{align}
Note the following properties of the rotation matrix:
 \begin{equation}
  \frac{\mathrm{d}}{\mathrm{d}\theta} R\big(\theta -\nicefrac{\pi}{2} \big) = R(\theta) \text{ and } 
  R( \theta - \nicefrac{\pi}{2} )=\begin{pmatrix}
        0 & 1\\
        -1 & 0
       \end{pmatrix}
 \end{equation}
 \begin{equation}
 R(\theta)=
 \begin{pmatrix}
  \cos(\theta) & -\sin(\theta) \\
  \sin(\theta) & \cos(\theta)
 \end{pmatrix}, \quad R(\theta)^{-1}= \begin{pmatrix}
  \cos(\theta) & \sin(\theta) \\
  -\sin(\theta) & \cos(\theta)
 \end{pmatrix} = R(-\theta).
\end{equation}
The exact solution of the characteristics in eqn.~\eqref{spectral:VP2d2v:split4:3} reads then
 \begin{align}
 \label{spectral:VP2d2v:scovel:charx}
  X(t)&= X(0) + \int_0^t \dot{X}(\tau) \,\mathrm{d}\tau
  = X(0) + \int_0^t V(\tau) \,\mathrm{d}\tau\\
  &= X(0) + \int_0^t R\big(-  \tau \frac{q }{m } B_3( X(0),0 ) \big) V(0) \,\mathrm{d}\tau\\
       &=X(0) + \frac{m }{q_s B_3 } \left[ R\big(- \theta + \nicefrac{\pi}{2} \big) - R(0) \right] V(0) \\
& = X(0)+ \frac{m }{q_s B_3 }\begin{pmatrix}
     \sin(\theta) & 1-\cos(\theta)  \\
  \cos(\theta)-1 &     \sin(\theta) 
\end{pmatrix} 
\begin{pmatrix}   V_1(0) \\ V_2(0)\end{pmatrix},
\\
 \label{spectral:VP2d2v:scovel:charv}
  V(t) &=  R(-\theta)   V(0) ,   \text{ for }  \theta= t \, \frac{q }{m } B_3( X(0),0 ).
 \end{align}
If we suppose $B_3$ to be constant then the advection in velocity space \eqref{spectral:VP2d2v:scovel:charv} is independent of
eqn.~\eqref{spectral:VP2d2v:scovel:charx}. Hence it is straightforward so solve \eqref{spectral:VP2d2v:scovel:charx} first and \eqref{spectral:VP2d2v:scovel:charv} thereafter.
By spatial Fourier transform the advection in $x$ can be integrated exactly, which yields
\begin{equation}
\label{spectral:VP2d2v:scovel:charx:discrete}
     \hat{f}(k_{x_1},k_{x_2},v,t) = \hat{f}(k_{x_1},k_{x_2},v,0) \expbb{ -   \frac{\imagi \, m }{q_s B_3 } k_x  \cdot  \begin{pmatrix}
      \sin( \theta) v_1 + ( 1- \cos(\theta)  ) v_2\\
       (\cos(\theta) -1) v_1 + \sin(\theta) v_2 \end{pmatrix} }\\ 
\end{equation}
Since only one dimensional Fourier transforms are used and the entire problem is two-dimensional for each $x$, aliasing can be suppressed by zero padding at
small costs compared to padding the entire distribution function.
The $v\times B$ rotation in \eqref{spectral:VP2d2v:scovel:charv} is the same as in eqn.~\eqref{spectral:VP2d2v:scovel:charv} and hence can be discretized as before e.g. by Fourier transform in $v$ with rotation by shearing.
For symmetric composition the adjoint method is needed, which means the rotation in $v$ has to be applied before the rotation and advection in $x$. To account for the fact, that the rotation \eqref{spectral:VP2d2v:scovel:charv}
is applied first, we rewrite eqn.~\eqref{spectral:VP2d2v:scovel:charv} and \eqref{spectral:VP2d2v:scovel:charx} into:
 \begin{align}
 \label{spectral:VP2d2v:scovel:adj:charv}
 V(0) &=  R(-\theta)^{-1} V(t) =R(\theta) V(t)  ,   \text{ for }  \theta= t \frac{q }{m } B_3( X(0),0 ),\\
 \label{spectral:VP2d2v:scovel:adj:charx}
  X(t) &=  X(0) + \int_0^t V(\tau) \,\mathrm{d}\tau\\
  &= X(0) + \int_0^t R\big(- \tau \frac{q }{m } B_3( X(0),0 ) \big) \underbrace{ V(0)}_{= R(\theta) V(t)} \,\mathrm{d}\tau\\
       &=X(0) + \frac{m }{q_s B_3 } \left[ R\big( -\theta + \nicefrac{\pi}{2} \big) - R(0) \right] R(\theta) V(t)\\ 
       &=X(0) + \frac{m }{q_s B_3 }       
       \begin{pmatrix}
     \sin(\theta) & 1-\cos(\theta)  \\
  \cos(\theta)-1 &     \sin(\theta) 
\end{pmatrix} 
       \begin{pmatrix}
     \cos(\theta) & -\sin(\theta)  \\
  \sin(\theta) &     \cos(\theta) 
\end{pmatrix} \begin{pmatrix}   V_1(0) \\ V_2(0)\end{pmatrix}\\
       &=X(0) +  \frac{m }{q_s B_3 } \begin{pmatrix}
     \sin(\theta) & \cos(\theta)-1  \\
  1-\cos(\theta)  &     \sin(\theta) 
\end{pmatrix}  \begin{pmatrix}   V_1(0) \\ V_2(0)\end{pmatrix}. 
 \end{align}
The discrete counterpart of \eqref{spectral:VP2d2v:scovel:adj:charx} reads then:
\begin{equation}
\label{spectral:VP2d2v:scovel:charx:adj:discrete}
    \hat{f}(k_{x_1},k_{x_2},v,t) = \hat{f}(k_{x_1},k_{x_2},v,0) \expbb{ -   \frac{\imagi\,m }{q_s B_3 } k_x  \cdot  \begin{pmatrix}
      \sin( \theta) v_1 + ( \cos(\theta) -1 ) v_2\\
       ( 1-\cos(\theta)) v_1 + \sin(\theta) v_2 \end{pmatrix} }.\\ 
\end{equation}
Since applying the adjoint after a forward time step with a negative time step ($\varphi(\Delta t) \varphi^*(-\Delta t) = Id$) corresponds exactly to the identity map we obtain a symmetric method by combining the Scovel and the adjoint Scovel.
For the discrete rotation by shearing in $v$ this is obvious, since $R(\theta) R(-\theta) = \mathbb{I}$. By combining \eqref{spectral:VP2d2v:scovel:charx:discrete}
and \eqref{spectral:VP2d2v:scovel:charx:adj:discrete} the in time symmetry is also easily verified:
\begin{multline}
 \expbb{ -   \frac{\imagi \, m }{q_s B_3 } k_x  \cdot  \begin{pmatrix}
      \sin( \theta) v_1 + ( 1- \cos(\theta) ) v_2\\
       (\cos(\theta)-1 ) v_1 + \sin(\theta) v_2 \end{pmatrix} } \cdot \\
 \expbb{ -   \frac{\imagi \,m }{q_s B_3 } k_x  \cdot  \begin{pmatrix}
      \sin( -\theta) v_1 + (  \cos(-\theta) -1 ) v_2\\
       ( 1- \cos(-\theta) ) v_1 + \sin(-\theta) v_2 \end{pmatrix} }\\
       =       \expbb{ -   \frac{\imagi \,m }{q_s B_3 } k_x  \cdot  \begin{pmatrix}
          \sin( \theta) v_1 + ( 1- \cos(\theta)  ) v_2 -\sin( \theta) v_1 + (  \cos(\theta)-1  ) v_2\\
       (\cos(\theta)-1 ) v_1 + \sin(\theta) v_2  + (1-\cos(\theta) ) v_1 - \sin(\theta) v_2 \end{pmatrix} }\\
       =       \expbb{ -   \frac{\imagi \,m }{q_s B_3 } k_x  \cdot  \begin{pmatrix} 0 \\ 0  \end{pmatrix} }
       =1
\end{multline}
Note that in the case of a constant homogeneous magnetic field the symmetrically composed Scovel
coincides with the splitting presented in~\cite{EINKEMMER20142144} and the symmetric methods (28) and (29) in \cite{knapp2015splitting}.\\
\subsection{Extension to \Ampere}
In the case of a homogeneous Maxwellian background one can solve the \Ampere~instead of the Poisson equation in order to obtain an update on the fields. In Fourier space
this reads
\begin{equation}
\label{spectral:ampere2d:fourier}
\hat{E}_j (k_x,t)=\hat{E}_j (k_x,0) + \int_0^t \iint v_j \hat{f}(k_x,v,\tau) \, \mathrm{d} (v_1,v_2)\, \mathrm{d}\tau, \quad j=1,2.
\end{equation}
Now any split step containing an advection in $x$ has to update the electric field according to eqn.~\eqref{spectral:ampere2d:fourier}.
For the exponential Boris the only relevant split step is
\begin{equation}
\begin{cases}
  \dot{V}(t) &= 0\\
\dot{X}(t)&= V(t) \\
  \end{cases} 
\end{equation}
such that the \Ampere~update for $j=1,2$, following eqn.~\eqref{spectral:vlasovampere:ampere1:fourier}, reads
\begin{multline} 
 \hat{E}_j (k_x,t)
 =\hat{E}_j (k_x,0) - \frac{q}{m} \int_0^t \iint v_j \hat{f}(k_x,v,0) \expbb{ -\imagi  ( \underbrace{v_1 k_{x_1} + v_2 k_{x_2}}_{=k_x\cdot v}) \tau }   \, \mathrm{d} (v_1,v_2)\, \mathrm{d}\tau \\
 =\hat{E}_j (k_x,0) - \frac{q}{m}  \begin{cases}
    t \iint v_j \hat{f}(k_x,v,0)  \, \mathrm{d} (v_1,v_2)                         &\text{for } k_x =0,\\
    \iint  \frac{v_j}{-   \imagi  (k_x \cdot v)        }  \left[ \expbb{ - \imagi t   (k_x \cdot v)   } -1 \right]   \hat{f}(k_x,v,0)   \, \mathrm{d} (v_1,v_2)   & \text{otherwise}.
  \end{cases}.
\end{multline}
By this technique Gauss' law is satisfied at any time. We recall that the electric field is obtained from the Poisson equation at any time as
\begin{equation}
 \hat{E}_j(k_x,t) =  \frac{ \imagi\, k_{x_j}}{k_x\cdot k_x}  q\int \hat{f}(k_x,v,t) \dv.
\end{equation}
For Scovel's method it is not as straightforward, such that this shall be treated elsewhere.

\subsection{Kelvin Helmholtz Instability}
\begin{figure}
 \centering
 \begin{tabular}{c  c}
    $||B_3||=1$ (weak) &  $||B_3||=32$ (strong) \\
    \hline \noalign{\vskip 2mm}    
        \multicolumn{2}{c}{electrostatic energies}\\ 
  \includegraphics[width=0.48\textwidth]{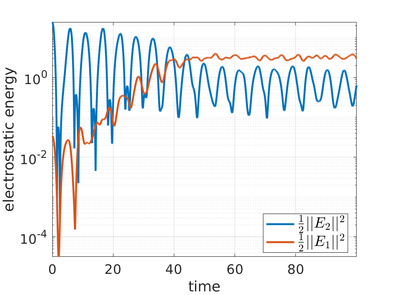} &
  \includegraphics[width=0.48\textwidth]{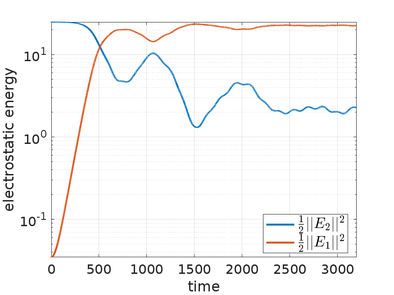}\\
   \hline \noalign{\vskip 2mm}    
        \multicolumn{2}{c}{kinetic energies}\\ 
  \includegraphics[width=0.48\textwidth]{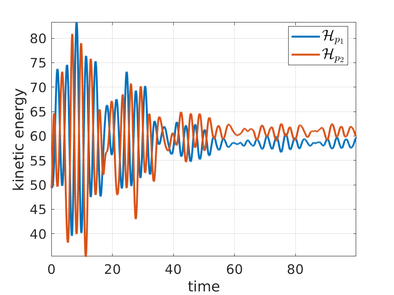} &
  \includegraphics[width=0.48\textwidth]{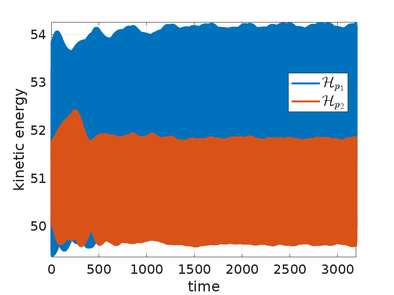}\\
\hline \noalign{\vskip 2mm}    
        \multicolumn{2}{c}{relative energy error}\\ 
  \includegraphics[width=0.48\textwidth]{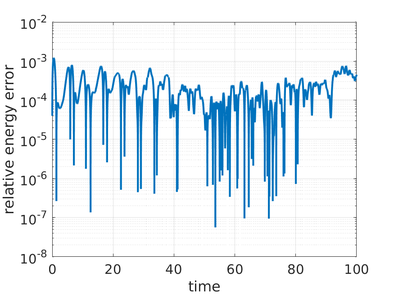} &
  \includegraphics[width=0.48\textwidth]{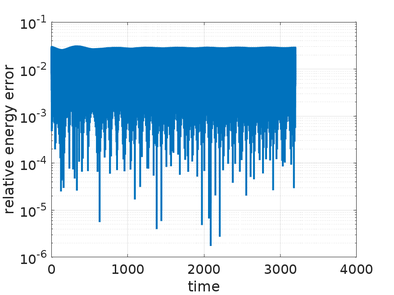}\\  
  \hline
  $\Delta t = 0.01, ~t_{\max}=100$ &  $\Delta t = 0.32, ~t_{\max}=3200$\\
  \hline
    \end{tabular}
\caption{Energies in the Kelvin Helmholtz instability under weak and strong magnetic field for $N_x=N_v=32,~\Delta t^{\psi} = 0.01, ~t^{\psi}_{\max}=100$ with Scovel's splitting.
         In both cases the stable mode $E_2$ looses energy with in the transition to the turbulent phase, where the system is driven by the unstabe mode $E_1$.
         Although the dynamics are highly nonlinear the energy error remains constant.}
\label{fig:khi:32:scovel:diag}
\end{figure}

\begin{figure}
\centering
 \begin{tabular}{c c}
    $||B_3||=1$ (weak) &  $||B_3||=32$ (strong) \\
    \hline \noalign{\vskip 2mm}    
     \multicolumn{2}{c}{$t^{\psi}=18$}\\
  \includegraphics[width=0.48\textwidth]{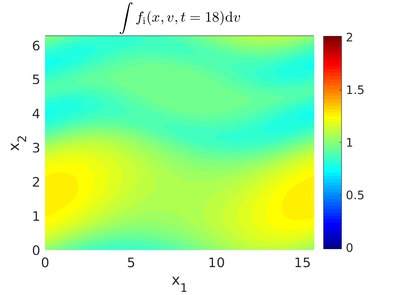} &
  \includegraphics[width=0.48\textwidth]{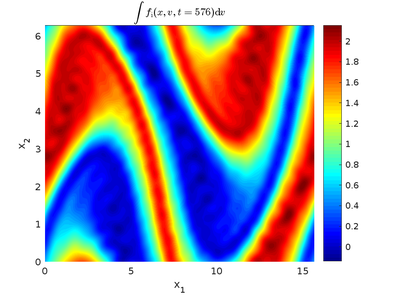}\\
  \hline \noalign{\vskip 2mm}    
     \multicolumn{2}{c}{$t^{\psi}=58$}\\
  \includegraphics[width=0.48\textwidth]{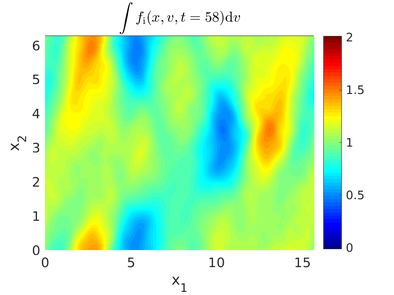} &
  \includegraphics[width=0.48\textwidth]{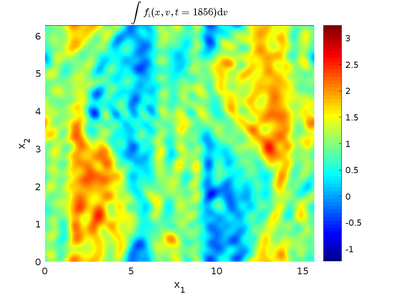}\\
  \hline \noalign{\vskip 2mm}    
     \multicolumn{2}{c}{$t^{\psi}=98$}\\
   \includegraphics[width=0.48\textwidth]{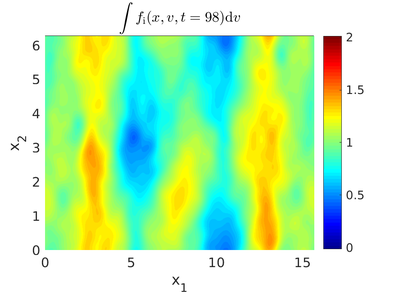} &
  \includegraphics[width=0.48\textwidth]{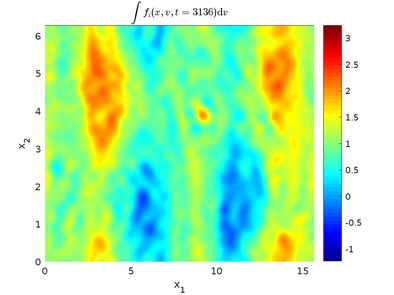}\\
  \hline
 \end{tabular}
\caption{Projection of the phase space onto the spatial plane. For the weak case the initial mode structure is less pronounced as there is less confinement. The lack of confinement appears to introduce
         diffusion like effects. Nevertheless turbulence evolves in both cases at later times.}
 \label{fig:khi:32:scovel:visu:fdv}
\end{figure}

\begin{figure}
\centering
 \begin{tabular}{c c}
        $||B_3||=1$ (weak) &  $||B_3||=32$ (strong) \\
      \hline \noalign{\vskip 2mm}    
     \multicolumn{2}{c}{$t^{\psi}=18$}\\
  \includegraphics[width=0.48\textwidth]{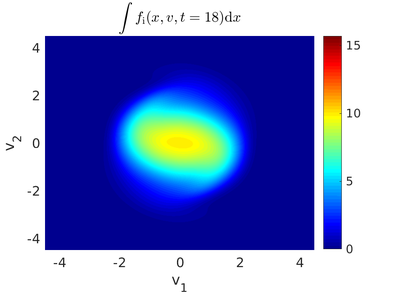} &
  \includegraphics[width=0.48\textwidth]{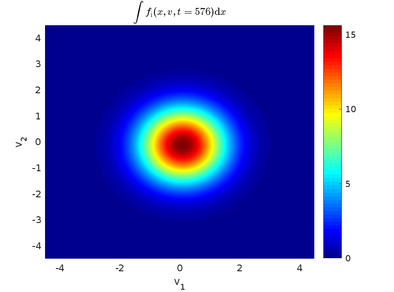}\\
  \hline \noalign{\vskip 2mm}    
     \multicolumn{2}{c}{$t^{\psi}=58$}\\
  \includegraphics[width=0.48\textwidth]{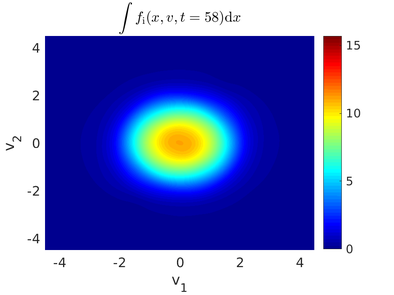} &
  \includegraphics[width=0.48\textwidth]{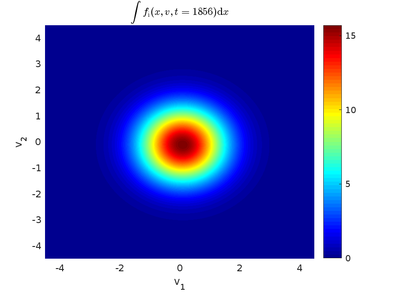}\\
  \hline \noalign{\vskip 2mm}    
     \multicolumn{2}{c}{$t^{\psi}=98$}\\
   \includegraphics[width=0.48\textwidth]{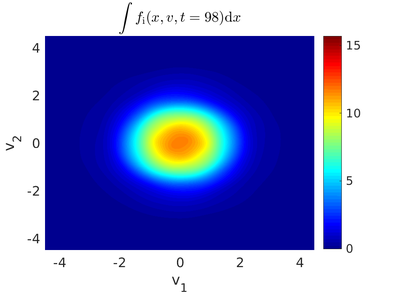} &
  \includegraphics[width=0.48\textwidth]{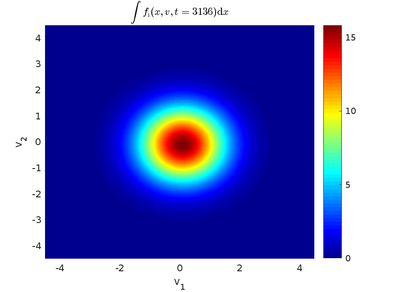}\\
  \hline
 \end{tabular}
 \caption{Projection of the phase space onto the velocity plane. Due to the lack of confinement by the weak magnetic field
          kinetic effects smear out the Maxwellian. For the strong field the distribution resembles a sharp Maxwellian, such that a fluid model
          based on precisely that assumption describes the dynamics very well and kinetic effects are not essential.}
 \label{fig:khi:32:scovel:visu:fdx}
\end{figure}

\begin{figure}
 \begin{tabular}{c c}
        $||B_3||=1$ (weak) &  $||B_3||=32$ (strong) \\
    \hline \noalign{\vskip 2mm}    
     \multicolumn{2}{c}{$t^{\psi}=18$}\\
  \includegraphics[width=0.48\textwidth]{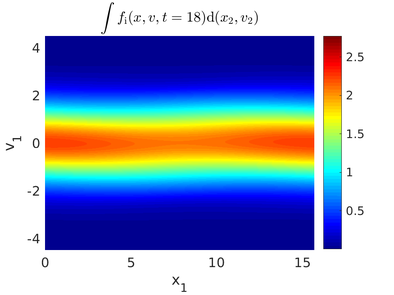} &
  \includegraphics[width=0.48\textwidth]{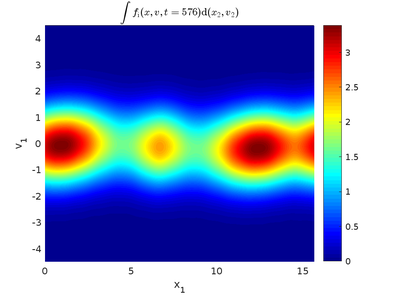}\\
  \hline \noalign{\vskip 2mm}    
     \multicolumn{2}{c}{$t^{\psi}=58$}\\
  \includegraphics[width=0.48\textwidth]{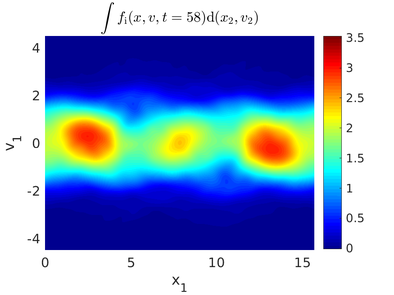} &
  \includegraphics[width=0.48\textwidth]{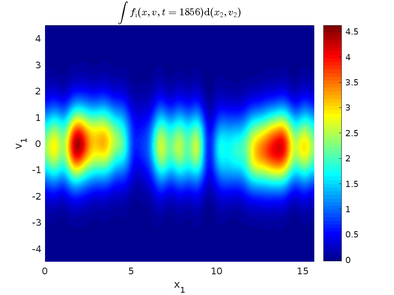}\\
  \hline \noalign{\vskip 2mm}    
     \multicolumn{2}{c}{$t^{\psi}=98$}\\
   \includegraphics[width=0.48\textwidth]{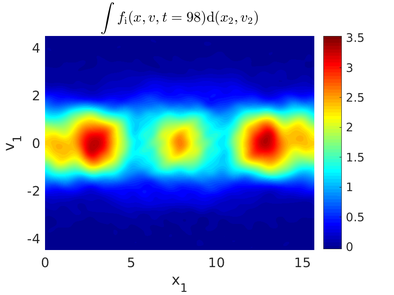} &
  \includegraphics[width=0.48\textwidth]{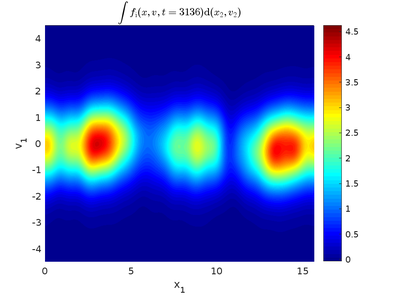}\\
  \hline
 \end{tabular}
 \caption{Projection of the phase space onto the $(x_1,v_1)$ plane in order to observe the kinetic structure of the unstable mode. For the strong magnetic field
         a pronounced turbulence in fig.~\ref{fig:khi:32:scovel:visu:fdv} is observed. Here this leads to much finer mode structure in the reduced phase space for the strong case compared to the weak one.}
 \label{fig:khi:32:scovel:visu:fdx1v1}
\end{figure}

We consider a two-dimensional periodic domain with the lengths
$L_d= \nicefrac{2\pi}{k_d}, ~d=1,2$ and the initial condition for the electrons $(q_{\mathrm{e}}=-1, ~ m_{\mathrm{e}}=1)$
\begin{equation}
 f_e(x_1,x_2, v_1,v_2,t=0) =  \left(1+ \sin(k_2 x_2)+ \nu \cos(k_1 x_1) \right) \frac{1}{2\pi} \exp\Big( \frac{ v_1^2+ v_2^2    }{2} \Big)
\end{equation}
along with a constant ion background $\rho_{\mathrm{i}}=1,~q_{\mathrm{i}}=1$ in the Poisson equation. In case of a strong magnetic field $B_3$
the dynamics of the fully kinetic model described by the Vlasov equation is very well approximated by the corresponding fluid model, a scaled version of the vorticity equation
\begin{align}
\label{eqn:vorticity:vp2d2v:limit}
\frac{\partial_t \rho_{\mathrm{e}} (x,t) }{\lVert B_3 \rVert} + q_{\mathrm{i}} \left[ E_1(x,t) \partial_{x_2} \rho_{\mathrm{e}} (x,t) -  E_2(x,t) \partial_{x_2} \rho_{\mathrm{e}} (x,t) \right]=0\\
\end{align}
coupled to the same fields stemming from the Poisson equation
\begin{align}
-\Delta  \Phi(x,t)  &=  \rho_{\mathrm{i}} + \rho_{\mathrm{e}}(x,t),\\
\quad E(x,t) &= - \nabla \Phi(x,t),\\
\rho_{\mathrm{e}}(x,t) &= q_{\mathrm{e}} \int f_{\mathrm{e}}(x,v,t)~\mathrm{d}v.
\end{align}
The detailed scalings and techniques can be found in~\cite{golse1999vlasov,frenod2000long,filbet2016asymptotically,filbet2017asymptotically}.
Here eqn.~\eqref{eqn:vorticity:vp2d2v:limit} is set on the kinetic time scale, but by introducing the fluid time scale 
with $t^{\psi} =  \frac{t}{ \lVert B_3 \rVert}$ and a perturbed fluid density, the same as in \cite{shoucri1981two} by removing the constant background
$\psi(x, t^{\psi}  ) =  \left[ \rho_{\mathrm{i}} +\rho_{\mathrm{e}}(x,  \lVert B_3 \rVert t^\psi ) \right]$. This allows us to use the results
on the linear stability of the Kelvin Helmholtz instability derived in~\cite{shoucri1981two}. Depending on the wave number $k$ the growth rate on the
fluid time scale in a periodic domain is by using a Taylor expansion \textit{approximated} as
\begin{equation}
\label{eqn:vorticity:khi:dispersion}
 \omega^\psi =  \imagi 2  (1-k)k    %
\end{equation}
and on the kinetic time scale
\begin{equation}
\label{eqn:vorticity:khi:dispersion:kinetic}
 \omega =  \frac{\imagi 2  (1-k)k }{\lVert B_3 \rVert}.
\end{equation}
In~\cite{shoucri1981two} the neutrally stable mode in the periodic domain was found in agreement to eqn.~\eqref{eqn:vorticity:khi:dispersion} at $k=1$. Therefore,
we install a linearly stable mode in the second dimension by $k_2=1$ and excite a linearly unstable mode in the first one $k_1=0.4$ with small amplitude $\nu=0.015$ in order to observe a Kelvin-Helmholtz instability 
with growth rate $\omega = \imagi 0.480$. \\
By rescaling to the fluid time scale we are able to compare the three different schemes from weak to strong magnetic field whilst holding the actual number of time steps constant.
We focus on the unstable mode and there fore the electrostatic energy in the first dimension $\frac{1}{2}||E_1||^2$. The growth rate is only known for the limit $||B_3|| \rightarrow \infty$, hence we
do not expect agreement for small $||B_3||$. But fig.~\ref{fig:vp2d2v_khi_discussion} shows that we approach the fluid model with increasing $||B_3||$. For $||B_3||=1$
all integrators show the same performance, but in the case of the strong magnetic field Scovel's splitting is clearly better. For 
$||B_3||=1	6$ the standard and exponential Boris splitting fail entirely whereas Scovel's splitting remains unaffected.
For $||B_3||=32$ the exponential Boris reverts back to a lower frequency, which is a known effect from integrating particle trajectories~\cite{patacchini2008explicit}, hence the steeper growth rate. This demonstrates
that it is worthwhile to actually include the spatial rotation into the numerics.
\begin{figure}[htb]
\centering

\begin{tabular}{c c}
   $\frac{1}{2}||E_1||^2$ & rel. energy error \\
\hline \noalign{\vskip 2mm}    
    \multicolumn{2}{c}{$||B_3||=1$}\\
  \includegraphics[width=0.48\textwidth]{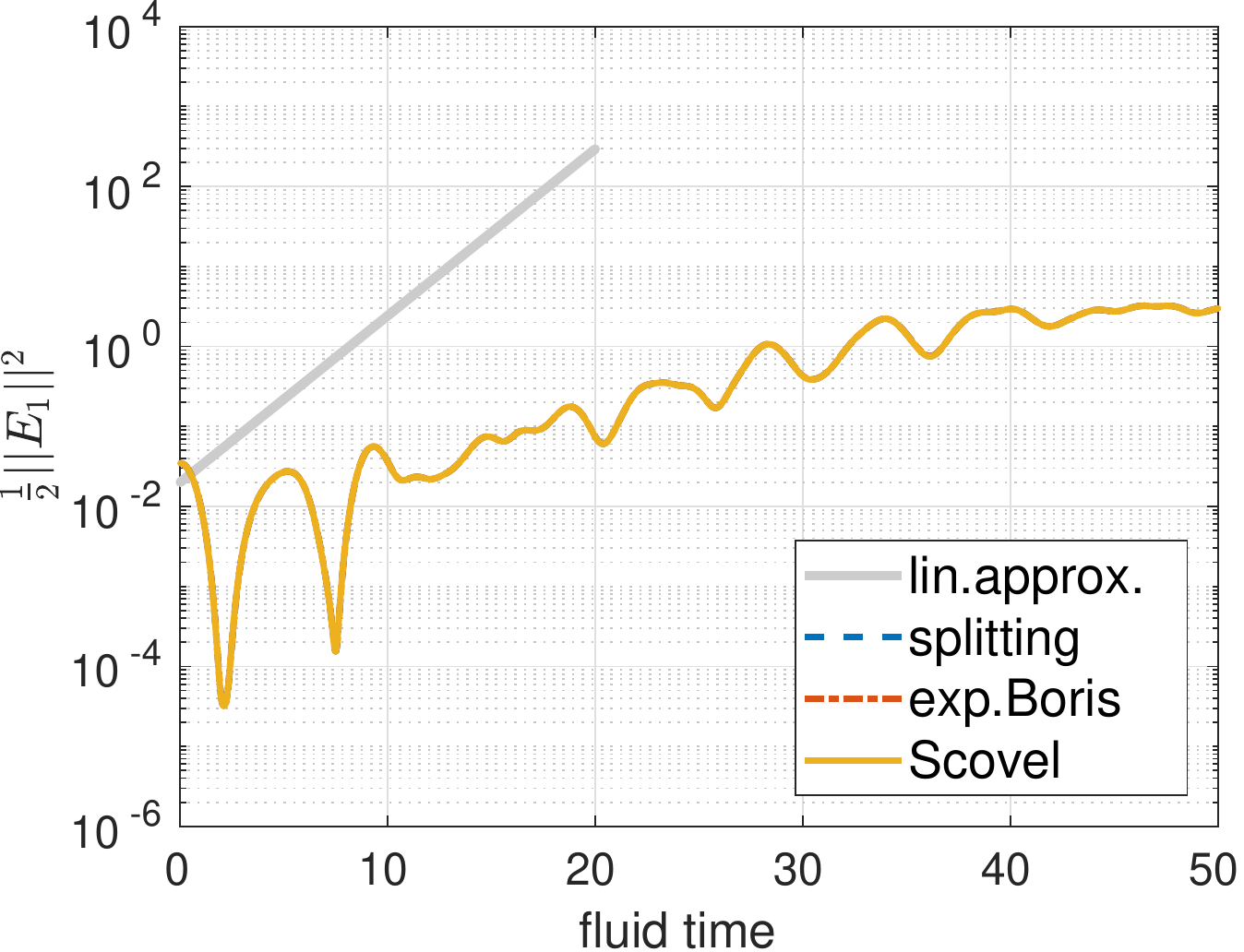}   & \includegraphics[width=0.48\textwidth]{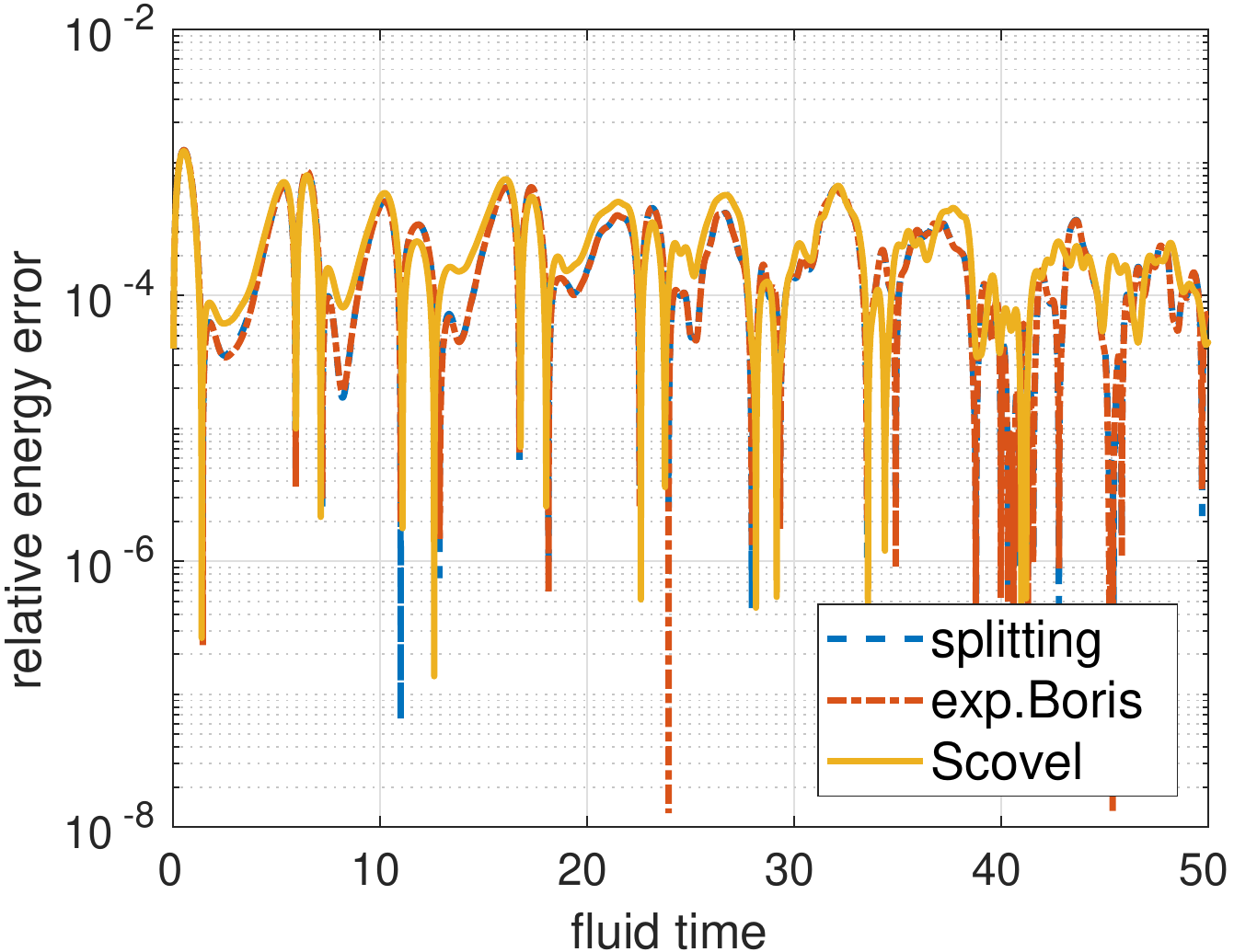} \\
  \hline \noalign{\vskip 2mm}    
\multicolumn{2}{c}{$||B_3||=2$}\\
   \includegraphics[width=0.48\textwidth]{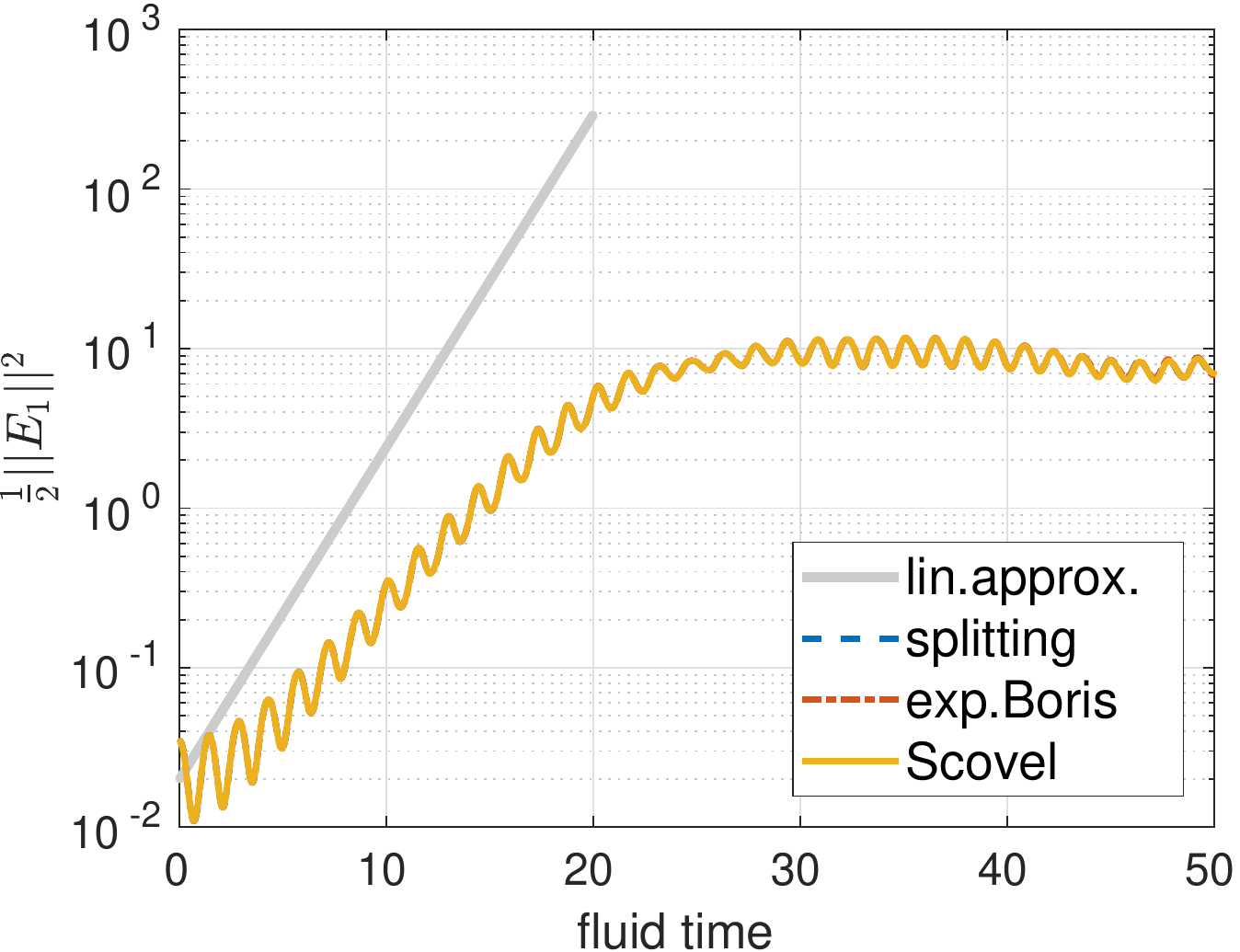}   & \includegraphics[width=0.48\textwidth]{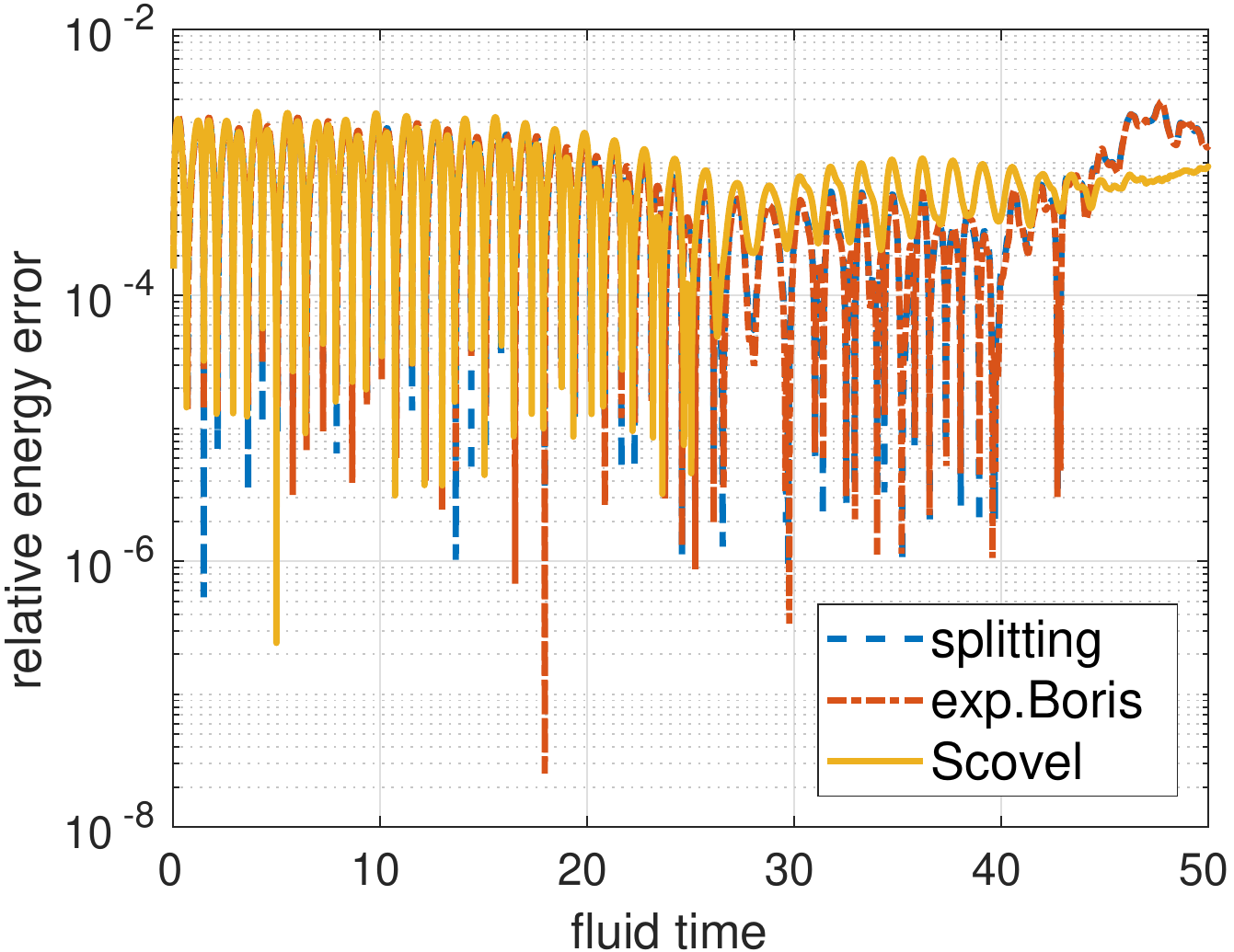} \\
   \hline \noalign{\vskip 2mm}    
\multicolumn{2}{c}{$||B_3||=4$}\\
  \includegraphics[width=0.48\textwidth]{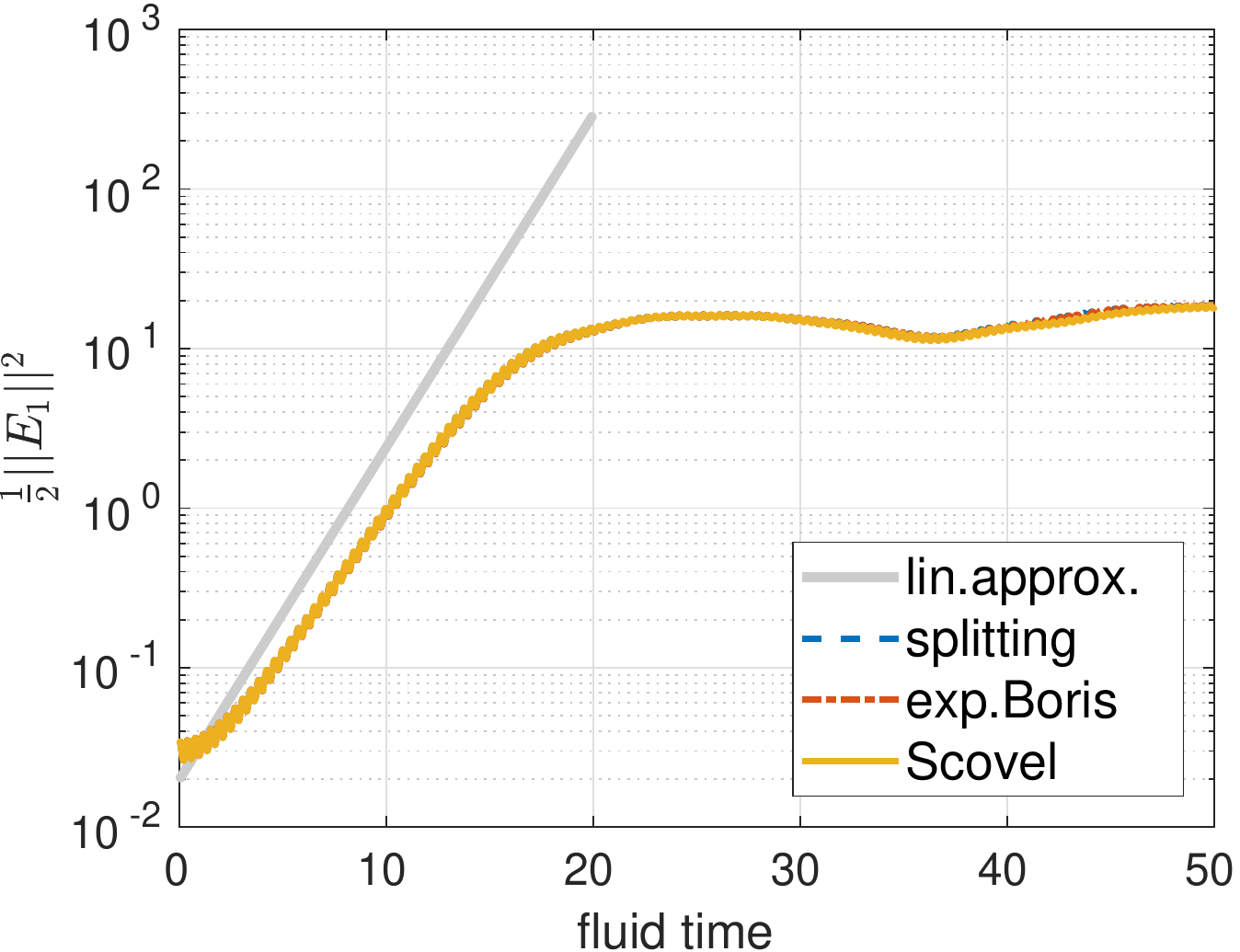}   & \includegraphics[width=0.48\textwidth]{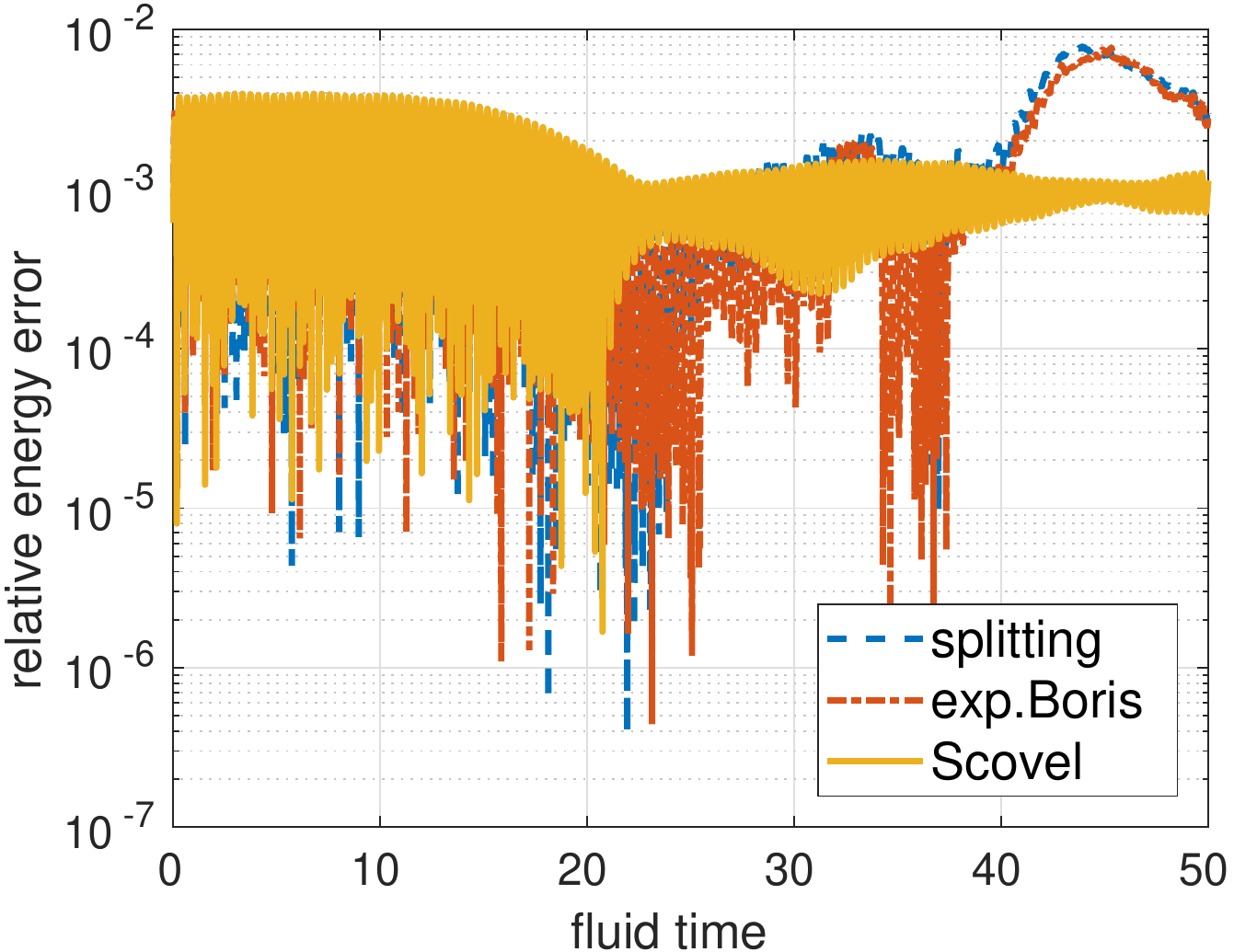} \\
   \hline
\end{tabular}

 \caption{Electrostatic energy of the unstable mode and relative energy error in the Kelvin Helmholtz instability for increasing magnetic field strength $||B||$. (See continuation)}
\end{figure}
\begin{figure}[htb]\ContinuedFloat
\begin{tabular}{c c}
$\frac{1}{2}||E_1||^2$ & rel. energy error \\
\hline \noalign{\vskip 2mm}    
\multicolumn{2}{c}{$||B_3||=8$}\\
  \includegraphics[width=0.48\textwidth]{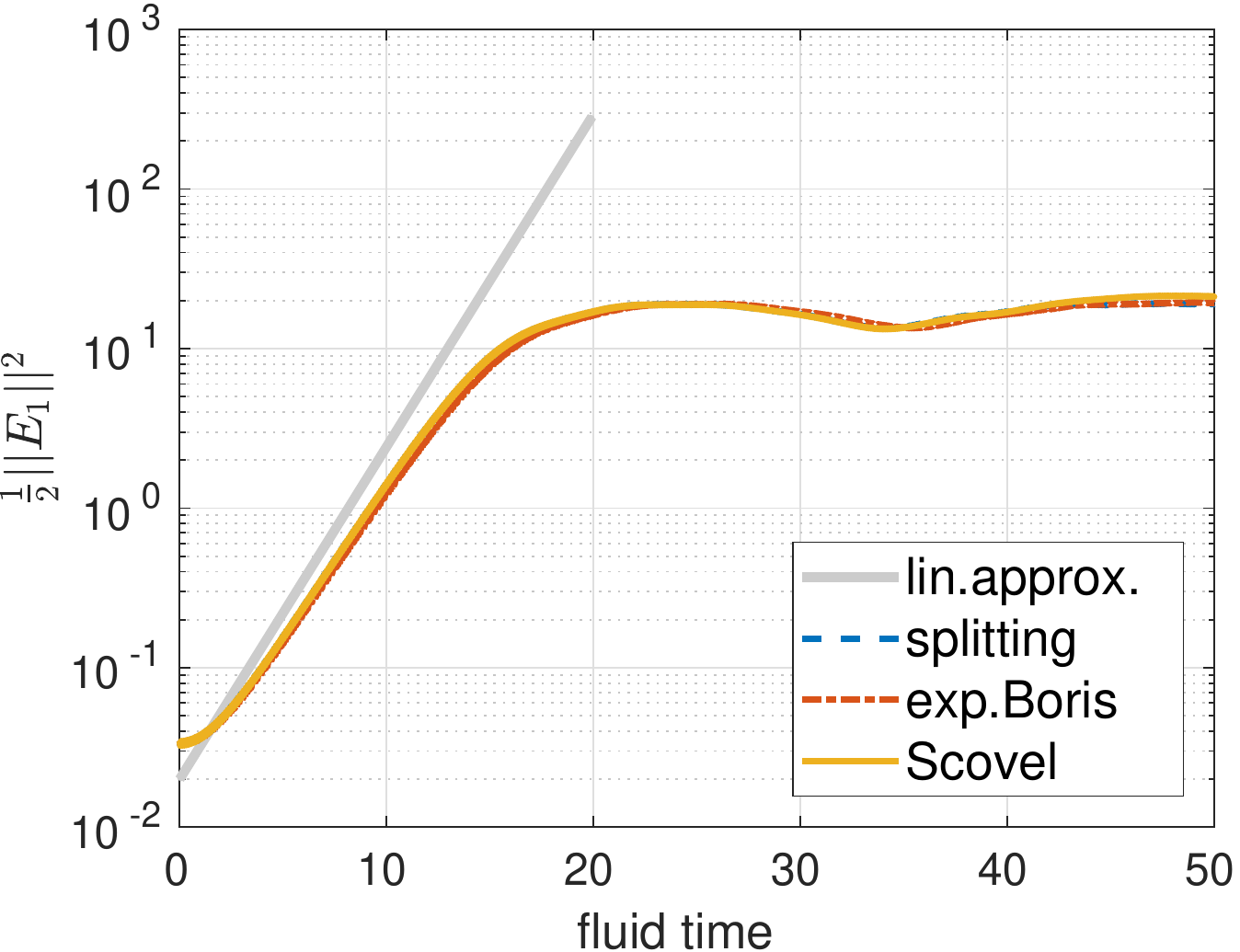}   & \includegraphics[width=0.48\textwidth]{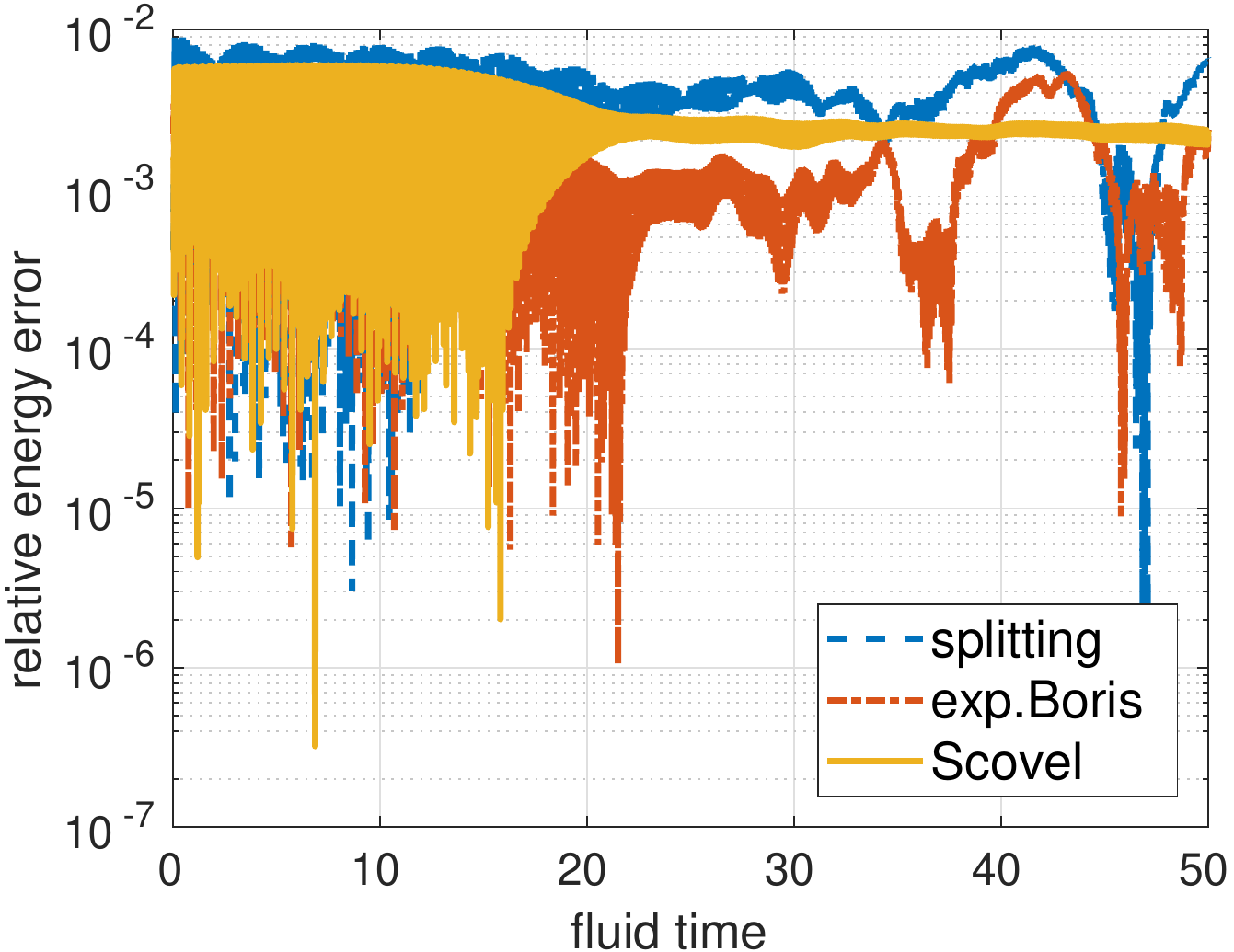} \\
  \hline \noalign{\vskip 2mm}    
\multicolumn{2}{c}{$||B_3||=16$}\\
   \includegraphics[width=0.48\textwidth]{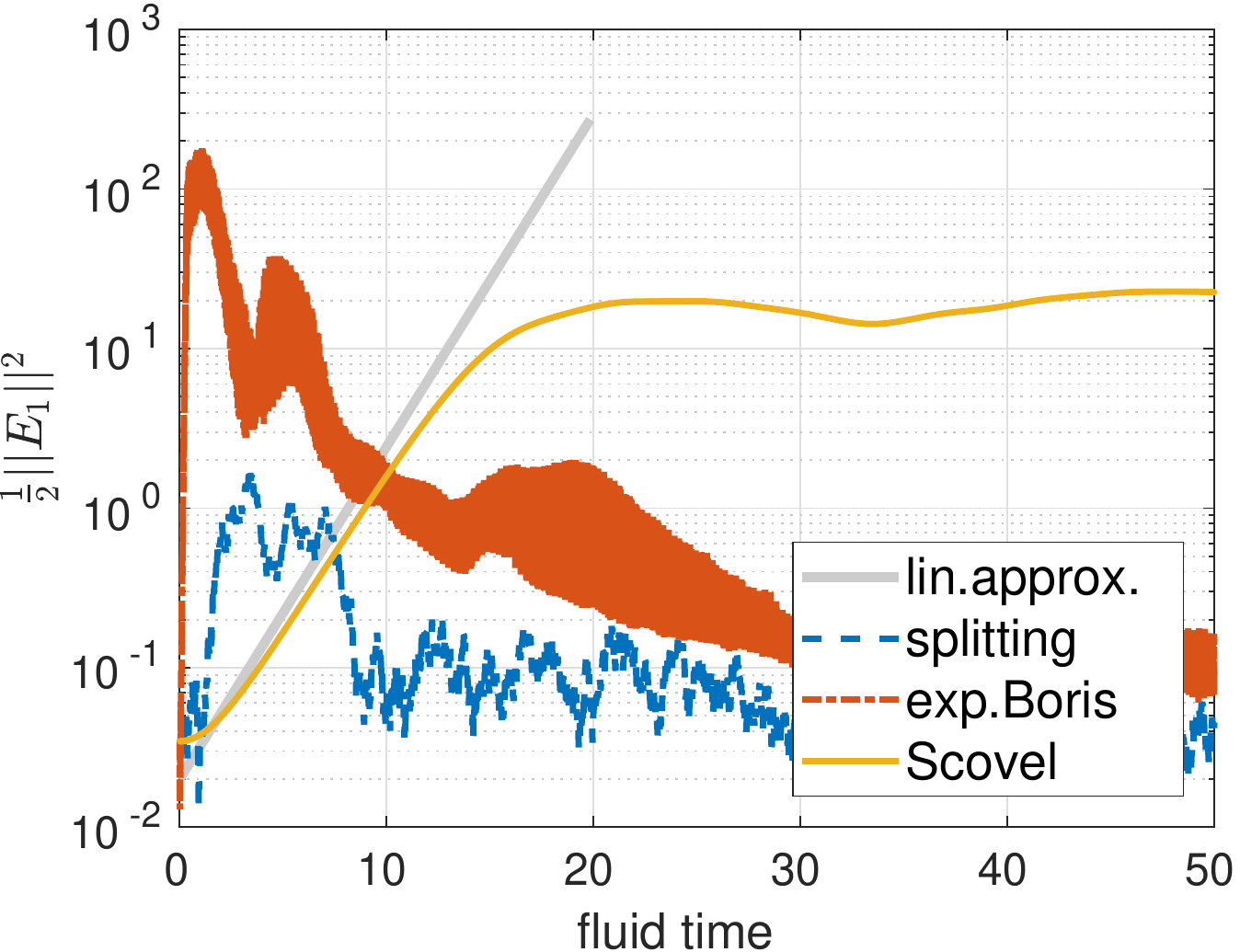}   & \includegraphics[width=0.48\textwidth]{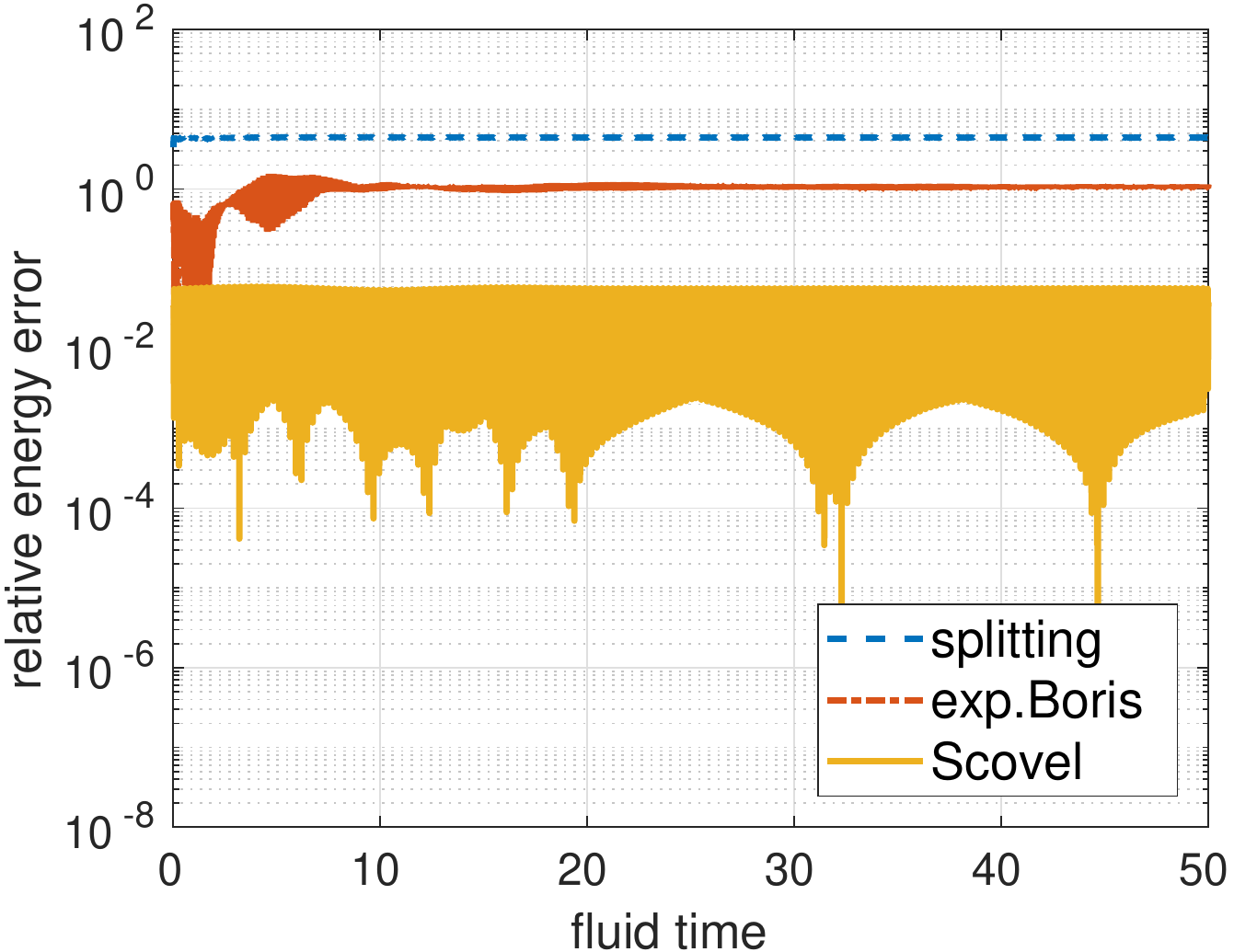} \\
   \hline \noalign{\vskip 2mm}    
\multicolumn{2}{c}{$||B_3||=32$}\\
  \includegraphics[width=0.48\textwidth]{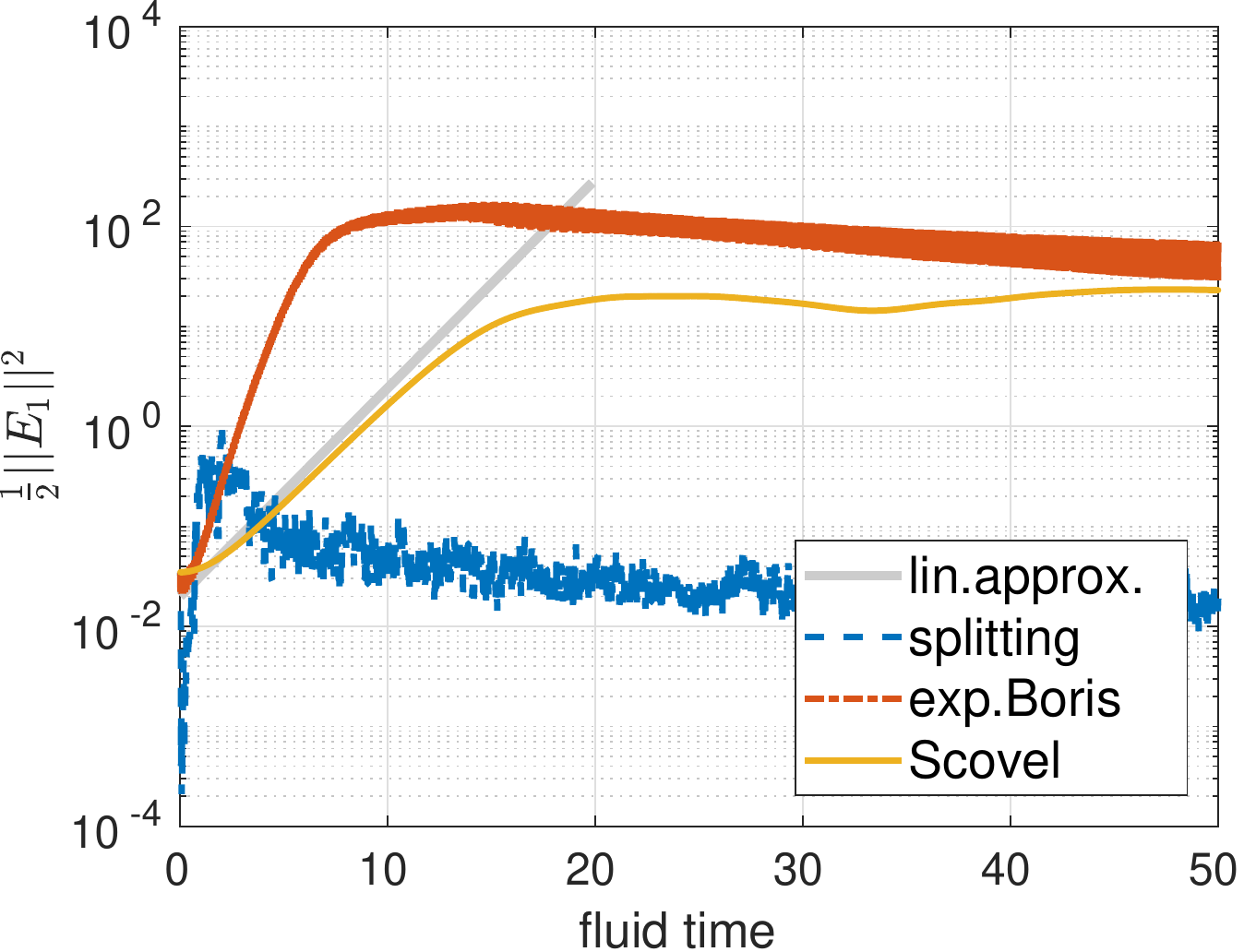}   & \includegraphics[width=0.48\textwidth]{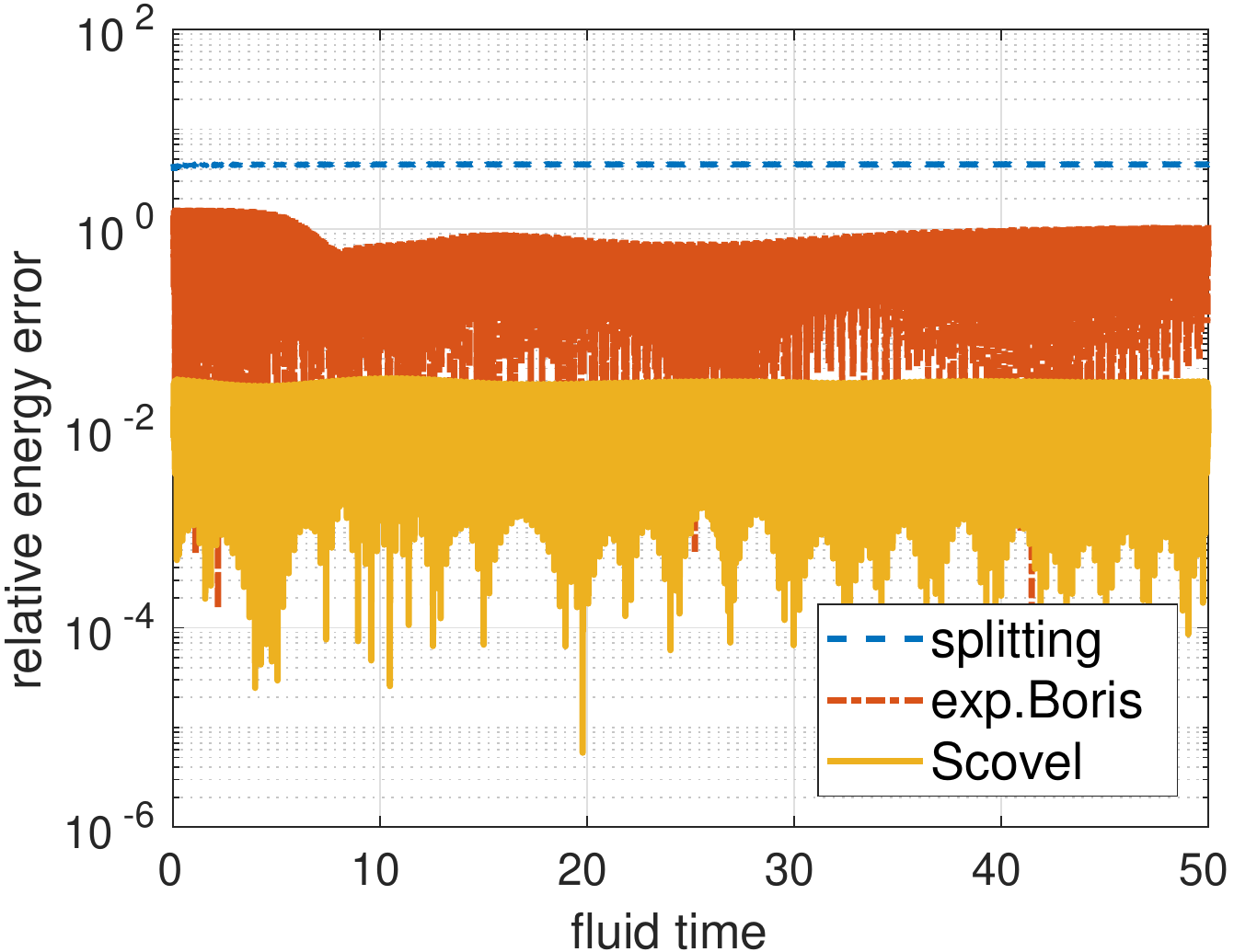} \\
  \hline
\end{tabular} 
\caption{Electrostatic energy of the unstable mode and relative energy error in the Kelvin Helmholtz instability for increasing magnetic field strength $||B||$, 
$N_x=32,~N_v=32,~\Delta t^{\psi} = 0.01$ and $t^{\psi}_{\max}=40 $. The actual number of time steps stays constant, which allows a comparison between the standard Strang splitting for the rotation.}
\label{fig:vp2d2v_khi_discussion}
\end{figure}
In the following we extend our investigation into the nonlinear phase using the superior Scovel method in case of a weak and strong field in
figures.~\ref{fig:khi:32:scovel:diag},~\ref{fig:khi:32:scovel:visu:fdx},~\ref{fig:khi:32:scovel:visu:fdx1v1} and \ref{fig:khi:32:scovel:visu:fdv}.
The entire system is driven by the unstable mode in $x1$, which deteriorates the stable mode in $x2$ leading to an energy loss in $E_2$ especially in the nonlinear phase.
This behavior is more pronounced in seems to be present for the stronger magnetic field. While the kinetic energy remains almost constant the only difference is the frequency of the
oscillation which is directly linked to the gyro-frequency.
Most importantly, fig.~\ref{fig:khi:32:scovel:diag} shows that the energy error remains despite the nonlinear dynamics constant over long time. It is also slightly higher in the case
of a strong field. This can be also seen in fig.~\ref{fig:khi:32:scovel:visu:fdv}, where the turbulence is much more pronounced for the strong field. The finer mode structure can also be seen
in fig.~\ref{fig:khi:32:scovel:visu:fdx1v1} explaining the higher electrostatic energy.
Fig.~\ref{fig:khi:32:scovel:visu:fdx} indicates already that kinetic effects are only present in the weak case and a fluid model based on a Maxwellian velocity distribution
is a fairly good approximation under a strong magnetic field.

\FloatBarrier
\section{Vlasov--Maxwell (1d2v)}
\begin{figure}
\begin{tabular}{c c c}
	\multicolumn{3}{c}{Low resolution: $N_x=N_{v_1}=N_{v_3}=32$, $\Delta t=0.05$}\\
\hline
electrostatic energy & energy error & momentum error\\
\hline
	\multicolumn{3}{c}{strong Landau damping $\mathcal{P}_{\epsilon}=4.17-14$}\\
  \includegraphics[width=0.31\textwidth]{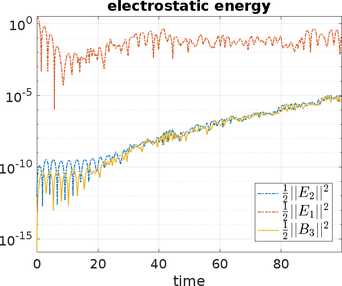}&
  \includegraphics[width=0.31\textwidth]{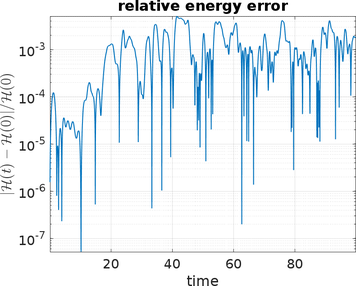}&
  \includegraphics[width=0.31\textwidth]{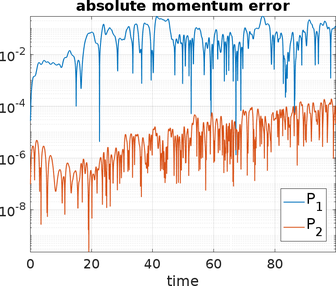}\\
\hline
	\multicolumn{3}{c}{Weibel instability $\mathcal{P}_{\epsilon}=2.9e-14$}\\
  \includegraphics[width=0.31\textwidth]{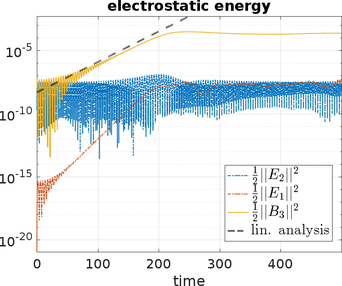}&
  \includegraphics[width=0.31\textwidth]{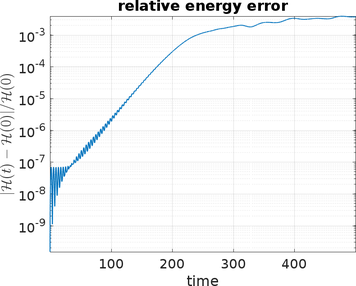}&
  \includegraphics[width=0.31\textwidth]{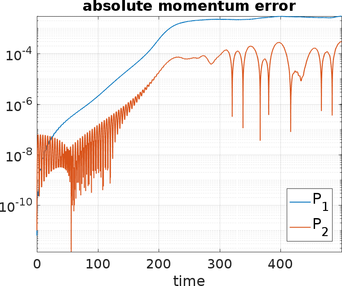}\\
\hline
\multicolumn{3}{c}{Weibel streaming instability (asym.) $\mathcal{P}_{\epsilon}=3.01e-13$}\\
\includegraphics[width=0.31\textwidth]{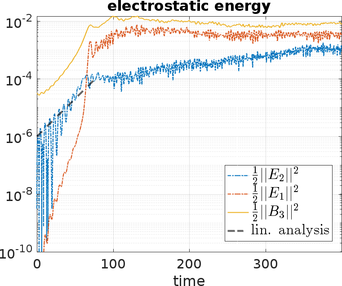}&
  \includegraphics[width=0.31\textwidth]{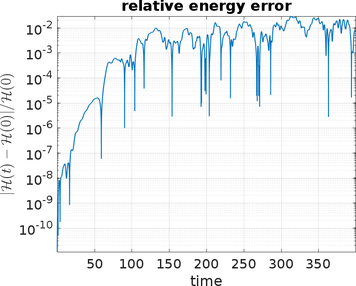}&
  \includegraphics[width=0.31\textwidth]{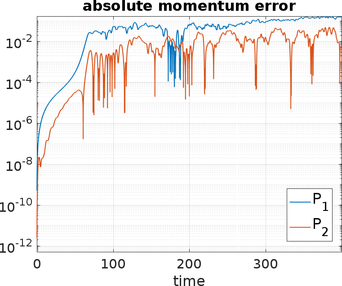}\\
\hline
\end{tabular}
\caption{Electrostatic energy, relative energy error and the momentum error
in the two velocity components for different test cases of the Vlasov--Maxwell 1d2v geometric pseudo-spectral solver. 
The time discretization is performed by a second order Strang splitting. Although the resolution with just $32$ grid points per dimension is very low,
the solver appears to be stable over longer times. }
\label{fig:spectral:vm1d2v:results_low}
\end{figure}

\begin{figure}
\begin{tabular}{c c c}
	\multicolumn{3}{c}{Low resolution: $N_x=N_{v_1}=N_{v_3}=32$, $\Delta t=0.05$}\\
\hline
	\multicolumn{3}{c}{strong Landau damping}\\
  \includegraphics[width=0.31\textwidth]{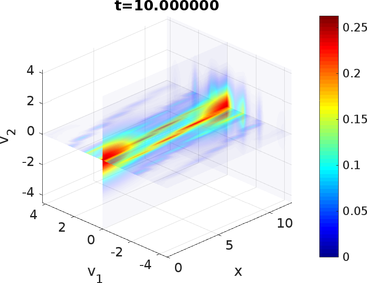}&
  \includegraphics[width=0.31\textwidth]{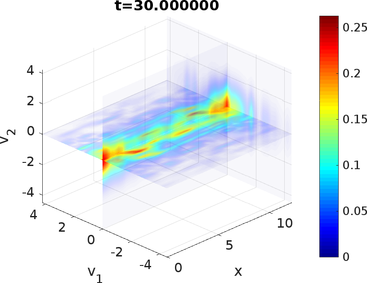}&
  \includegraphics[width=0.31\textwidth]{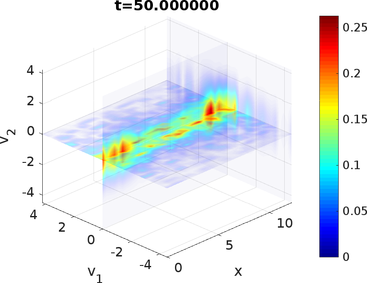}\\
\hline
	\multicolumn{3}{c}{Weibel instability}\\
  \includegraphics[width=0.31\textwidth]{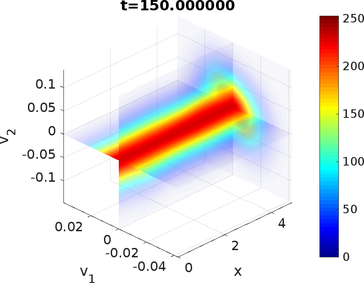}&
  \includegraphics[width=0.31\textwidth]{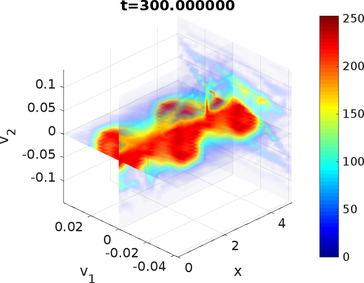}&
  \includegraphics[width=0.31\textwidth]{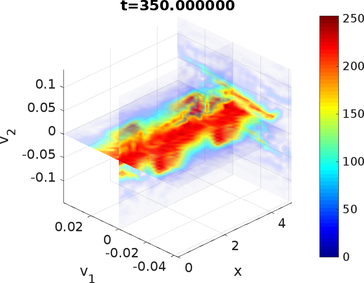}\\
 \hline
\multicolumn{3}{c}{Weibel streaming instability (asym.)}\\
  \includegraphics[width=0.31\textwidth]{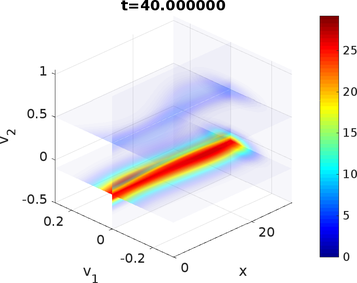}&
  \includegraphics[width=0.31\textwidth]{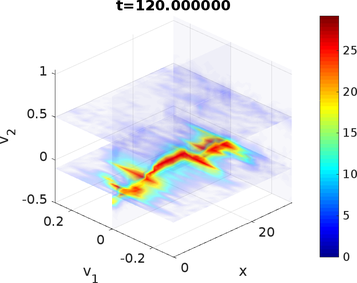}&
  \includegraphics[width=0.31\textwidth]{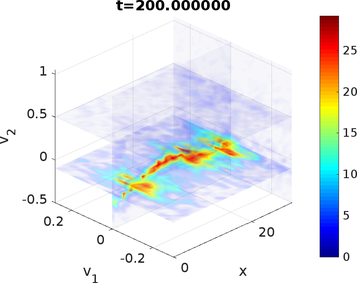}\\
\end{tabular}
\caption{Phase space densities for Vlasov--Maxwell 1d2v simulations under low resolution.}
\label{fig:spectral:vm1d2v:results_low2}
\end{figure}

\begin{figure}
\begin{tabular}{c c c}
	\multicolumn{3}{c}{High resolution: $N_x=N_{v_1}=N_{v_3}=128$, $\Delta t=0.01$}\\
\hline
electrostatic energy & energy error & momentum error\\
\hline
	\multicolumn{3}{c}{strong Landau damping $\mathcal{P}_{\epsilon}=5.97e-13$}\\
  \includegraphics[width=0.31\textwidth]{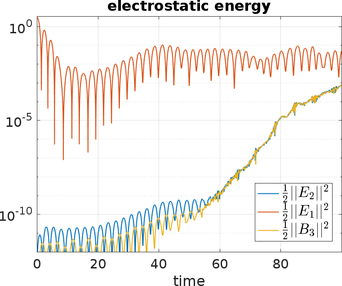}&
  \includegraphics[width=0.31\textwidth]{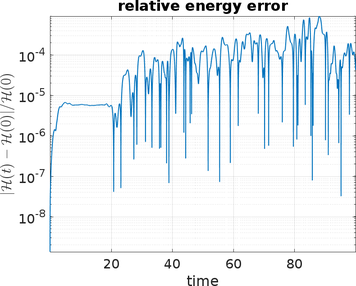}&
  \includegraphics[width=0.31\textwidth]{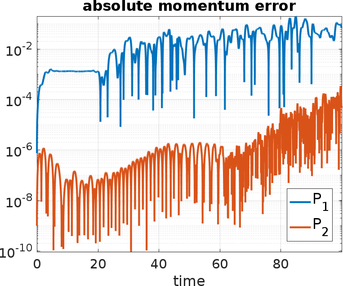}\\
\hline
\multicolumn{3}{c}{Weibel instability $\mathcal{P}_{\epsilon}=9.31e-14$}\\
  \includegraphics[width=0.31\textwidth]{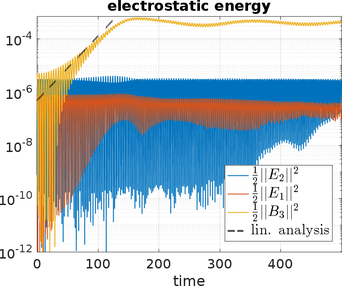}&
  \includegraphics[width=0.31\textwidth]{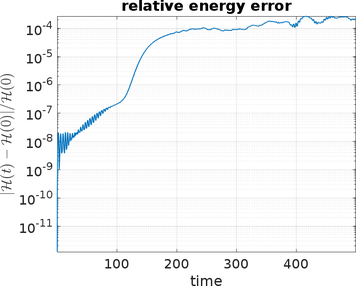}&
  \includegraphics[width=0.31\textwidth]{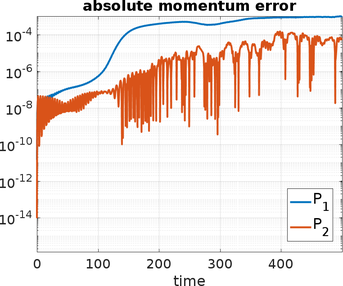}\\
\hline
\multicolumn{3}{c}{Weibel streaming instability (sym.) $\mathcal{P}_{\epsilon}=1.71e-12$}\\
\includegraphics[width=0.31\textwidth]{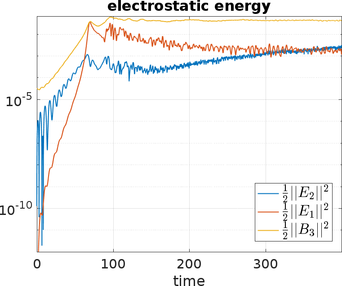}&
  \includegraphics[width=0.31\textwidth]{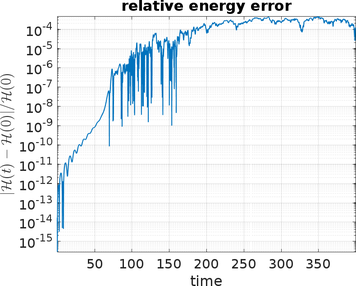}&
  \includegraphics[width=0.31\textwidth]{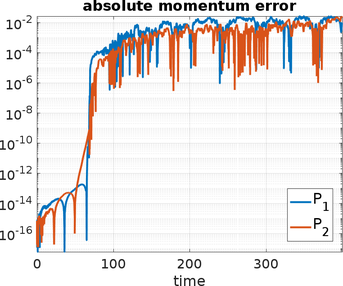}\\
\hline 
\multicolumn{3}{c}{Weibel streaming instability (asym.) $\mathcal{P}_{\epsilon}=7.8e-12$}\\
\includegraphics[width=0.31\textwidth]{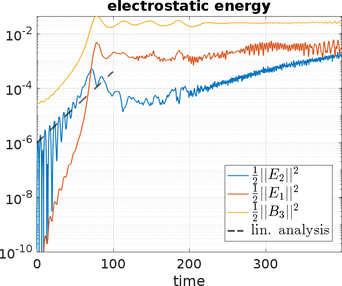}&
  \includegraphics[width=0.31\textwidth]{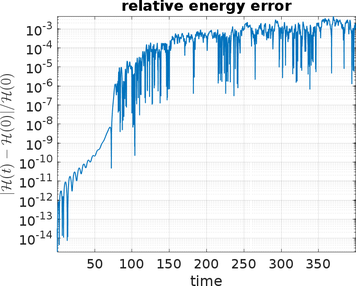}&
  \includegraphics[width=0.31\textwidth]{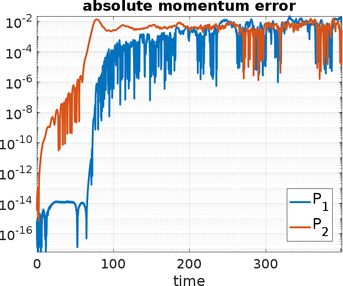}\\
 \hline
\end{tabular}
\caption{High resolution results for three Vlasov--Maxwell 1d2v simulations with the geometric pseudo-spectral solver.
       The energy error is smaller than in the low resolution but remains at a high level, which is comparable to the GEMPIC\cite{kraus2016gempic} results,
       where a smaller energy error was only achieved with a high order splitting.}
\label{fig:spectral:vm1d2v:results_high}
\end{figure}

\begin{figure}
\begin{tabular}{c c c}
	\multicolumn{3}{c}{High resolution: $N_x=N_{v_1}=N_{v_3}=128$, $\Delta t=0.01$}\\
\hline
	\multicolumn{3}{c}{strong Landau damping}\\
  \includegraphics[width=0.31\textwidth]{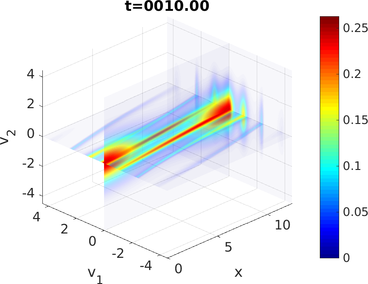}&
  \includegraphics[width=0.31\textwidth]{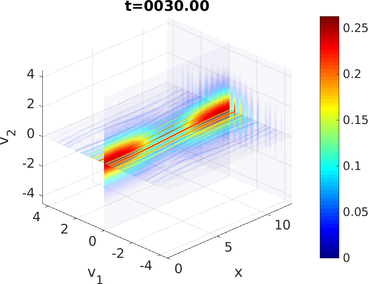}&
  \includegraphics[width=0.31\textwidth]{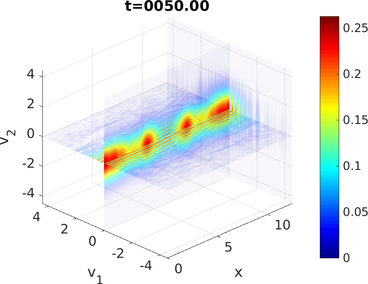}\\
\hline
\multicolumn{3}{c}{Weibel instability}\\
   \includegraphics[width=0.31\textwidth]{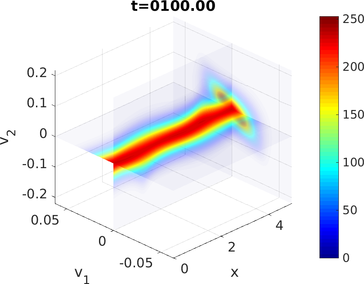}&
  \includegraphics[width=0.31\textwidth]{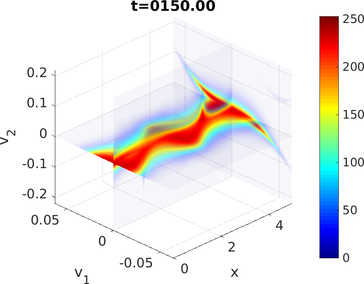}&
  \includegraphics[width=0.31\textwidth]{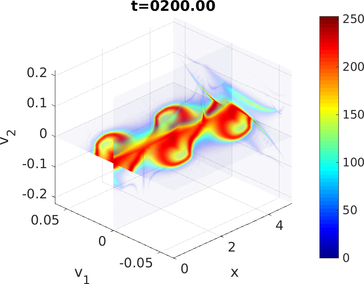}\\
    \includegraphics[width=0.31\textwidth]{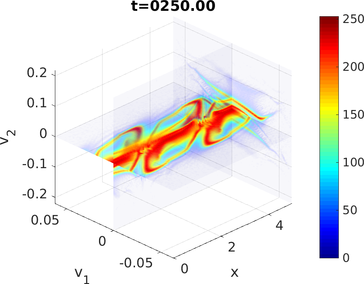}&
  \includegraphics[width=0.31\textwidth]{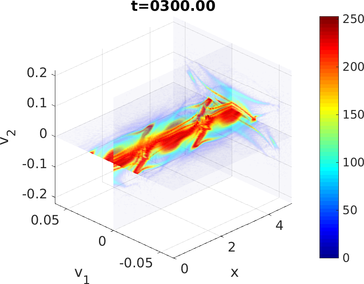}&
  \includegraphics[width=0.31\textwidth]{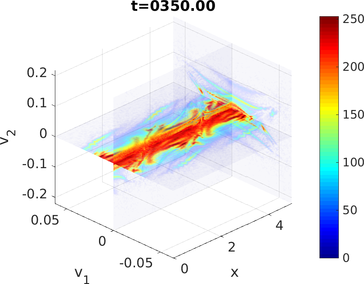}\\
   \hline
   \end{tabular}
\caption{Phase space densities for Vlasov--Maxwell 1d2v simulations under high resolution.}
\label{fig:spectral:vm1d2v:results_high2}
\end{figure}

\begin{figure}
\centering 
\begin{tabular}{cc}
   \includegraphics[width=0.35\textwidth]{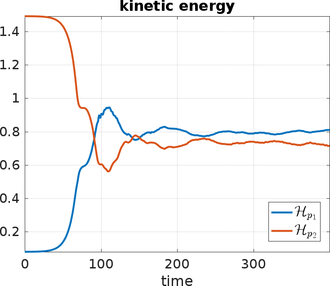}&
   \includegraphics[width=0.35\textwidth]{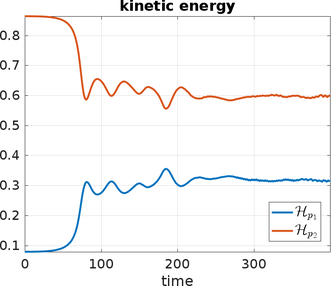}\\   
   symmetric & asymmetric \\
\end{tabular}
 \caption{Kinetic energy for the symmetric and asymmetric Weibel streaming instability at high resolution.}
 \label{fig:spectral:vm1d2v:results_high3}
\end{figure}

We consider a reduction of the full six-dimensional Vlasov--Maxwell model onto one spatial and two velocity components.
Elimination of the second and third spatial component, leaves us with two components of the electric field
and one component of the magnetic field. Here the single magnetic component in $z$-direction is denoted by $B_3$.
\begin{equation}
 x=x_1, ~v=(v_1,v_2), ~E=(E_1,E_2),~ B=B_3
\end{equation}
For a density $f(x,v_1,v_2,t)$, the two components of the electric field $E_1(x,t),E_2(x,t)$ and the 
magnetic field $B(x,t)$ the reduced Vlasov equation is given in eqn.~\eqref{vm:vlasov1d2v}.
\begin{equation}
\label{vm:vlasov1d2v}
\partial_t f_s + v_1 \partial_x f_s + \frac{q }{e} \frac{m_e}{m }
\left[  E_1 \partial_{v_1} f_s  +  E_2 \partial_{v_2} f_s 
 + B \left(  v_2 \partial_{v_1} f_s - v_1 \partial_{v_2} f_s \right)  \right] =0
\end{equation}
Dropping the species index $s$ yields the corresponding characteristics in eqn.~\eqref{vm:vlasov1d2v:char}. 
\begin{equation}
\label{vm:vlasov1d2v:char}
\begin{split}
 \frac{\mathrm{d}}{\mathrm{d}t} V_{1}(t) &= \frac{q}{e} \frac{m_e}{m}  \left[ E_1(X_s(t),t) + V_{2}(t) B(X(t),t) \right]\\
 \frac{\mathrm{d}}{\mathrm{d}t} V_{2}(t) &= \frac{q}{e} \frac{m_e}{m}  \left[ E_2(X_s(t),t) - V_1(t) B(X(t),t) \right]\\
 \frac{\mathrm{d}}{\mathrm{d}t} X(t) &= V_{1}(t)
 \end{split}
\end{equation}
The time dependent Maxwell equations reduce then to a system of three equations~\eqref{vm:maxwell1d2v}.
\begin{equation}
\begin{split}
\label{vm:maxwell1d2v}
 \partial_t E_1(x,t) &= - \sum_s \frac{q }{\mathrm{e}} \int v_1  f_s(x,v_1,v_2,t) \dv \\
 \partial_t E_2(x,t) &= -\left(\frac{c}{v_{th,e} }\right)^2 \partial_x B(x,t) - \sum_s \frac{q }{e} \int v_2  f_s(x,v_1,v_2,t) \dv,\\
 \partial_t B(x,t) &= - \partial_x E_2(x,t) 
\end{split}
\end{equation}
At the initialization for $t=0$ the Poisson eqn.~\eqref{vm1d2v:poisson1d} needs to be solved in order
to obtain the first component $E_1$ of the electric field. The second component is always initialized as zero, $E_2(x,0)=0$.
\begin{equation}
\label{vm1d2v:poisson1d}
 -\partial_{xx} \Phi(x,t) =  \sum_s \frac{q }{e}  \int_{\mathbb{R}^d}  f_s(x,v_1,v_2,t) \dv, \quad
 E_1(x,t)=-\partial_x \Phi(x,t)
\end{equation}
Here we chose $\frac{c}{v_{th,e}}=1$ and consider only the electrons $f=f_{\mathrm{e}},\, q=\mathrm{e} m=m_{\mathrm{e}}$, with a constant ion background $f_{\mathrm{i}}= \exp( - \frac{v_1^2+ v_2^2}{2})$.
The Hamiltonian splitting was already discussed extensively for Lagrangian particles~\cite{kraus2016gempic}, nevertheless, it is also possible to derive the same method for a
spectral discretization. For a different, but incorrect \cite{qin2015comment}, splitting this has already been done in \cite{crouseilles2015hamiltonian}.
Here we use the correct Hamiltonian splitting from~\cite{he2015hamiltonian}.
Let $f(x,v_1,v_2,t)$ denote the plasma density and $\hat{f}$ the Fourier transform. Since there are six different combinations of transforms
$\hat{f}$ denotes a transformation, where the transformed dimension is indicated as before by $k_x$, $k_{v_1}$ or $k_{v_2}$ in the argument. That means $\hat{f}(k_x, v_1, k_{v_2})$
denotes the Fourier transform of $f$ in $x$ and $v_2$. We begin by treating the Hamiltonian splitting for time integration from $0$ to $t$.
\begin{itemize}
 \item Kinetic energy $(d=1)$, $\mathcal{H}_{p_1}=\frac{1}{2}\iiint v_1^2 f(x,v,t) \,\mathrm{d}x \mathrm{d}v_1 \mathrm{d}v_2$ 
 \begin{equation}
 \label{spectralVM:Hp1}
  \begin{split}
    \partial_t f(x,v_1,v_2,t)   +  v_1 \partial_x f(x,v_1,v_2,t)  - \frac{q}{m} B_3(x,t) v_1 \partial_{v_2} f(x,v_1,v_2,t)  &= 0\\
  \partial_t B_3(x,t) &=0\\
  \partial_t E_1(x,t) = - q \int_{v_2^{\min}}^{v_2^{\max}} \int_{v_1^{\min}}^{v_1^{\max}} v_1  f(x,v_1,v_2,t)\, \mathrm{d} v_1 \mathrm{d}v_2
 \end{split}
 \end{equation}
The first problem, but luckily the only problem we will encounter, is the Fourier transform for the Vlasov density, since 
Fourier transforming in $x$ and $v_1$ simultaneously results in terms containing convolutions:
\begin{equation}
  \partial_t \hat{f}(k_x,v_1,k_{v_2},t)   +  v_1  \imagi k_x \hat{f}(k_x,v_1,k_{v_2},t) 
  - \frac{q}{m} \hat{B}_3(k_x,t) *_{k_x} v_1 \imagi k_{v_2} \hat{f}(k_x,v_1,k_{v_2},t)  = 0.
\end{equation}
This can be avoided by considering only the Fourier transform in $v_2$ such that \eqref{spectralVM:Hp1} can be solved exactly by
\begin{equation}
\begin{split}
\label{spectralVM:Hp1:odekv2}
  &\partial_t \hat{f}(x,v_1,k_{v_2},t)   +  v_1  \partial_x \hat{f}(x,v_1,k_{v_2},t) 
  - \frac{q}{m} B_3(x,0)  v_1 \imagi k_{v_2} \hat{f}(x,v_1,k_{v_2},t)  = 0\\
  \Leftrightarrow~ &\partial_t \hat{f}(x,v_1,k_{v_2},t) = - \left[  v_1  \partial_x ~
  - \frac{q}{m} B_3(x,0)  v_1 \imagi k_{v_2}  \right] \hat{f}(x,v_1,k_{v_2},t) \\
  \Rightarrow ~&   \hat{f}(x,v_1,k_{v_2},t) = \exp\left\{ - t \underbrace{ v_1 \left[  \partial_x ~
  - \frac{q}{m} B_3(x,0)  \imagi k_{v_2} \right]}_{=\mathcal{L}} \right\} \hat{f}(x,v_1,k_{v_2},0) 
\end{split}
\end{equation}
Here the exponential contains still the derivative $\partial_x$ which can be --- and this is a critical point here --- exactly obtained at the grid points $x_1, \dots x_{N_x}$
for the spectral discretization by Fourier forth and back-transform. For this recall that the discrete Fourier transform can be denoted in a matrix\footnote{Instead of assembling
the matrix by hand, one can just Fourier transform an identity matrix of the appropriate size.
In this way one always obtains the correct normalization, e.g. in MATLAB $fft(eye(N_x),[],1)$ and $ifft(eye(N_x),[],1)$.}
$\mathcal{F}_x \in \mathbb{R}^{N_x\times N_x}$ and $\mathcal{F}_x^{-1}$. Hence the matrix $L\in \mathbb{R}^{N_x \times N_x}$ 
representing the discrete but exact counterpart of $\mathcal{L}$ reads
\begin{equation}
 L= \underbrace{  v_1  \mathcal{F}_x^{-1} \mathrm{diag} \left(  \imagi k_1,\dots ,  \imagi k_{N_x} \right)\mathcal{F}_x}_{=L_A} 
 \underbrace{- v_1 \frac{q}{m} \imagi k_{v_2} \mathrm{diag} \left( B(x_1,0),\dots , B(x_{N_x},0)  \right)}_{L_L}.
\end{equation}
By calculating the matrix exponential $\exp(-tL)$ the systems of ODE arising from evaluating eqn.~\eqref{spectralVM:Hp1:odekv2} at every spatial grid point can be solved
exactly for each $v_1$ and $k_{v_2}$. Now it is obviously highly questionable to replace a fast Fourier transform by multiplication with a dense matrix, and although there are matrix free variants
of the standard algorithms available~\cite{al2011computing} we follow a much simpler approach. Note that $\exp(t L_A)$ and $\exp(t L_L)$ are as (transformed) diagonal matrices trivial to calculate
respectively to apply onto a vector $\left( \hat{f}(x_1,v_1,k_{v_2}),\dots, \hat{f}(x_{N_x},v_1,k_{v_2}) \right)$ but unfortunately $L_L$ and $L_A$ do not commute.
In such a situation Moler~\cite{moler2003nineteen} suggests to use the Trotter product formula
\begin{equation}
  \expbb{-\frac{t}{m} L} \expbb{-\frac{t}{m} (L_A+ L_L) }= \lim_{m \rightarrow \infty}
   \left( \expbb{-\frac{t}{m} L_A}  \expbb{-\frac{t}{m} L_L} \right)^m.
\end{equation}
Essentially this means, we should split  $ \hat{\mathcal{H}}_{p_1}$ into two parts which can be solved exactly in Fourier space and then sub-step these parts to the desired accuracy.
Splitting eqn. \eqref{spectralVM:Hp1} in the Vlasov--\Ampere~$\mathcal{H}_{p_{1,A}}$ part 
and the remaining terms of the Lorentz force $\mathcal{H}_{p_{1,L}}$ yields
\begin{equation}
  \begin{split}
  \mathcal{H}_{p_{1,A}}&
    \begin{cases}
     \partial_t f(x,v_1,v_2,t)   +  v_1 \partial_x f(x,v_1,v_2,t) =0,\\
  \partial_t E_1(x,t) = - q\int_{v_2^{\min}}^{v_2^{\max}} \int_{v_1^{\min}}^{v_1^{\max}} v_1  f(x,v_1,v_2,t) \,\mathrm{d} v_1 \mathrm{d}v_2,    
    \end{cases}    \\
   \mathcal{H}_{p_{1,L}}&
   \begin{cases}
    \partial_t f(x,v_1,v_2,t)   - \frac{q}{m} B_3(x,t) v_1 \partial_{v_2} f(x,v_1,v_2,t)  = 0,\\
  \partial_t B_3(x,t) =0.\\    
    \end{cases}
 \end{split}
 \end{equation}
The advection in $\mathcal{H}_{p_{1,A}}$ can be again directly solved by a Fourier transform in $x$,
 \begin{equation}
  \label{spectralVM:Hp1Af}
 \hat{f}(k_x,v_1,v_2,\tau) =\hat{f}(k_x,v_1,v_2,0)  \expbb{- v_1 \imagi k_x \tau} \text{ for } \tau \in[0,t].
 \end{equation}
The electric field is, identical as in Vlasov--Amp\`{e}re, obtained by inserting the time evolution \eqref{spectralVM:Hp1Af} yielding:
\begin{equation}
\label{spectralVM:Hp1AE}
 \begin{split}
   \hat{E}(k_x,t) &= 
   \hat{E}(k_x,0)  -  q\int_0^t \int_{v_2^{\min}}^{v_2^{\max}} \int_{v_1^{\min}}^{v_1^{\max}}  v_1\hat{f}(k_x,v_1,v_2,\tau) \, \mathrm{d}\tau  \mathrm{d} v_1 \mathrm{d}v_2\\
   &=\hat{E}(k_x,0)  -  q\int_0^t \int_{v_2^{\min}}^{v_2^{\max}} \int_{v_1^{\min}}^{v_1^{\max}}  v_1\hat{f}(k_x,v_1,v_2,0)  \expbb{- v_1 \imagi k_x \tau}\,   \mathrm{d}\tau  \mathrm{d} v_1 \mathrm{d}v_2\\
   &=\hat{E}(k_x,0) -  q\int_{v_2^{\min}}^{v_2^{\max}} \int_{v_1^{\min}}^{v_1^{\max}}  v_1\hat{f}(k_x,v_1,v_2,0)  \int_0^t \expbb{- v_1 \imagi k_x \tau} \, \mathrm{d}\tau  \mathrm{d} v_1 \mathrm{d}v_2 \\
   &=\hat{E}(k_x,0) -
  q \int_{v_2^{\min}}^{v_2^{\max}} \int_{v_1^{\min}}^{v_1^{\max}}  v_1\hat{f}(k_x,v_1,v_2,0)  \frac{1}{- v_1 \imagi k_x } \left[      \expbb{- v_1 \imagi k_x \tau}   \right]_0^t \,\mathrm{d} v_1 \mathrm{d}v_2\\
 &= \hat{E}(k_x,0) 
  +  q \int_{v_2^{\min}}^{v_2^{\max}} \int_{v_1^{\min}}^{v_1^{\max}}   \hat{f}(k_x,v_1,v_2,0)\frac{1}{ \imagi k_x }\left[ \expbb{- v_1 \imagi k_x t}  -1       \right]\, \mathrm{d} v_1 \mathrm{d}v_2.
 \end{split}
\end{equation}
The second part $\mathcal{H}_{p_{1,L}}$ reduces to a constant coefficient advection in $v_2$ and is solved directly by
\begin{equation}
  \label{spectralVM:Hp1Lf}
 \hat{\mathcal{H}}_{p_{1,L}} \Big\{~ \hat{f}(x,k_{v_2},t)=    \expbb{\frac{q}{m} B_3(x,t) v_1~ \imagi k_{v_2} t}   \hat{f}(x,k_{v_2},0).
\end{equation}
Note that the split step $\mathcal{H}_{p_{1,A}}$, given in eqns.~\eqref{spectralVM:Hp1Af} and~\eqref{spectralVM:Hp1AE} can also be performed in $v_2$ transformed space,
thus, both eqn.\eqref{spectralVM:Hp1:dx} and eqn.~\eqref{spectralVM:Hp1:dxdv2} can be used.
\begin{equation}
\label{spectralVM:Hp1:dx}
 \hat{\mathcal{H}}_{p_{1,A}} \begin{cases}
 \hat{f}(k_x,v_1,v_2,t) &=\hat{f}(k_x,v_1,v_2,0)  \expbb{- v_1 \imagi k_x t}\\
   \hat{E}(k_x,t) &= \hat{E}(k_x,0)  +  q\int_{v_2^{\min}}^{v_2^{\max}} \int_{v_1^{\min}}^{v_1^{\max}}   \hat{f}(k_x,v_1,v_2,0)\frac{1}{ \imagi k_x }\left[ \expbb{- v_1 \imagi k_x t}  -1       \right]\, \mathrm{d} v_1 \mathrm{d}v_2.
  \end{cases}
\end{equation}
\begin{equation}
\label{spectralVM:Hp1:dxdv2}
 \hat{\mathcal{H}}_{p_{1,A}} \begin{cases}
 \hat{f}(k_x,v_1,k_{v_2},t) =\hat{f}(k_x,v_1,k_{v_2},0)  \expbb{- v_1 \imagi k_x t}\\
   \hat{E}(k_x,t) = \hat{E}(k_x,0)  \\
   \quad \qquad\qquad+~q \int_{v_1^{\min}}^{v_1^{\max}}  
   \hat{f}(k_x,v_1,k_{v_2}=0,0)\frac{1}{ \imagi k_x }\left[ \expbb{- v_1 \imagi k_x t}  -1 \right] \,\mathrm{d} v_1 \left( v_2^{\max} - v_2^{\min} \right)
  \end{cases}
\end{equation}
In order to obtain a symmetric splitting of $\mathcal{H}_{p_1}$ the following two second order options are available by Strang splitting, where $\varphi$ denotes the corresponding flux:
\begin{equation}
\begin{split}
 \varphi_{p_1} ( \Delta t) &=  \varphi_{p_{1,A}}\left(\frac{\Delta t}{2}\right) \circ \varphi_{p_{1,L}} ( \Delta t) 
                      \circ \varphi_{p_{1,A}}\left( \frac{\Delta t}{2}\right)\\
\varphi_{p_1} ( \Delta t) &=  \varphi_{p_{1,L}}\left(\frac{\Delta t}{2}\right) \circ \varphi_{p_{1,A}} ( \Delta t) 
                      \circ \varphi_{p_{1,L}}\left( \frac{\Delta t}{2}\right)\\
\end{split}
\end{equation}
With and without sub-stepping of this sub-splitting there was no visible difference (relative error at $\sim 10^{-6}$ to the fields obtained with the exact full matrix exponential for our test-cases, although
there is a difference to the exact integration, see fig.~\ref{vm1d2vspectral:weibels:Hp1_approx}. For the sake of efficiency we used only the single split step in the presented simulations. The reason for this could be that
the advection in eqn.~\eqref{spectralVM:Hp1Lf} takes only place in the $v_2$-component such that it would not affect
the integration of the Amp\`{e}re eqn.~\eqref{spectralVM:Hp1AE} in $\mathcal{H}_{p_{1,A}}$ where the velocity $v_2$ is integrated out.
This means that the resulting field $E$ is exactly the same as in the original $\mathcal{H}_{p_1}$ and Gauss' law is conserved.
\item Kinetic energy $(d=2)$,  $ \mathcal{H}_{p_2}=\frac{1}{2}\iiint v_2^2 f(x,v,t) \,\mathrm{d}x \mathrm{d}v_1 \mathrm{d}v_2$
  \begin{equation}
  \begin{split}
  \partial_t f(x,v_1,v_2,t) +  \frac{q}{m} v_2 B_3(x,t) \partial_{v_1} f(x,v_1,v_2,t) &=0 \\
  \partial_t E_2(x,t) = - q \int_{v_2^{\min}}^{v_2^{\max}} \int_{v_1^{\min}}^{v_1^{\max}} v_2  f(x,v_1,v_2,t) \,\mathrm{d} v_1 \mathrm{d}v_2
  \end{split}
 \end{equation}
 Since there is no advection in $x$ we know that the transport in $v_1$ averages out by
\begin{equation}
  \int_{v_1^{\min}}^{v_1^{\max}} f(x,v_1,v_2,\tau)\,  \mathrm{d} v_1 = \int_{v_1^{\min}}^{v_1^{\max}} f(x,v_1,v_2,0)  \mathrm{d} v_1  \quad \forall  \tau \in [0,t],
\end{equation}
such that $ \mathcal{H}_{p_2}$ can be integrated exactly in a single step yielding the final discretization
\begin{equation}
 \hat{\mathcal{H}}_{p_2} ~
\begin{cases}
\hat{f}(x,k_{v_1},v_2,t ) = 
\hat{f}(x,k_{v_1},v_2, 0 )  \expbb{-\imagi k_{v_1}  v_2 \frac{q}{m}B_3(x,0) t  }\\
\hat{E}_2(k_x,t)= \hat{E}_2(k_x, 0) - t \cdot q\int_{v_2^{\min}}^{v_2^{\max}} \int_{v_1^{\min}}^{v_1^{\max}}  v_2 \hat{f}(k_x, v_1,v_2, 0 ) \,\mathrm{d} v_1 \mathrm{d}v_2.
\end{cases} 
\end{equation}
 \item Electric energy, $\mathcal{H}_{E}= \frac{1}{2}\int \lvert E(x,t)\rVert^2\, \mathrm{d} x $
 \begin{equation}
 \begin{split}
   \partial_t f +  \frac{q}{m} E_1(x,t) \partial_{v_1}f(x,v_1,v_2,t) +  \frac{q}{m} E_2(x,t) \partial_{v_2}f(x,v_1,v_2,t) &=0\\
 \partial_t B_3(x,t) &= - \partial_x E_2(x,t)\\
 \partial_t E(x,t)& =0 
 \end{split}
 \end{equation}
 The advection is constant in $(v_1,v_2)$ and varies only in $x$, such that the constant coefficient advection can be solved exactly in Fourier space.
 \begin{equation}
 \hat{\mathcal{H}}_{E} ~
 \begin{cases}
 \hat{f}(x,k_{v_1},k_{v_2},t) =\hat{f}(x,k_{v_1},k_{v_2},0)
  \expbb{- \imagi \frac{q}{m} \left( E_1(x,0)k_{v_1} + E_2(x,0)k_{v_2} \right) t}\\
  \hat{B}_3(k_x,t) = \hat{B}_3(k_x,0)-  t \cdot \imagi k_x \hat{E}_2(k_x,t)\\
 \end{cases}
 \end{equation}
\item Magnetic energy, $\mathcal{H}_{B}= \frac{1}{2}\int \lVert B(x,t)^2\rVert\, \mathrm{d} x$
\begin{equation}
\begin{split}
  \partial_t E_2(x,t) &=- \partial_x B_3(x,t)\\
  \partial_t E_1(x,t) &= \partial_t B_3(x,t) =0
  \end{split}
 \end{equation}
 \begin{equation}  
   \hat{\mathcal{H}}_{B} ~\Big \{ \hat{E}_2(k_x,t) = \hat{E}_2(k_x,0) - t~ \imagi k_{x} \hat{B}(k_x,0)
 \end{equation}
\end{itemize}
For the initialization of the simulation the electric field $E_1$ is obtained by the Poisson equation, which reduces in one dimension to Gauss' law.
In Fourier space Gauss' law reads
\begin{equation}
\label{spectralVM:Gauss}
 \hat{E}_1(k_x,t) = \frac{1}{\imagi k_x}  \underbrace{q\int_{v_2^{\min}}^{v_2^{\max}} \int_{v_1^{\min}}^{v_1^{\max}}  \hat{f}(k_x, v_1,v_2, t )\, \mathrm{d} v_1 \mathrm{d}v_2}_{:=\hat{\rho}(k_x,t)} ~
 \text{ for } k_x \neq 0.
\end{equation}
Gauss' law is preserved during the entire simulation, such that we denote the error on eqn.~\eqref{spectralVM:Gauss}
at final time as $\mathcal{P}_{\epsilon}$, which should be close to machine precision.
Instead of the standard second order Strang splitting using two Lie steps, we prefer a second order method which has less than half the error constant of the Strang splitting \cite{mclachlan1995numerical}.
It requires four Lie steps and is given by symmetric composition of a flux $\varphi$ with its adjoint $\varphi^*$ as
\begin{equation}
\label{spectral:symmsplit2ndlie}
\begin{split}
 \varphi_{ \alpha \Delta t } \circ  \varphi_{ (\nicefrac{1}{2}- \alpha) \Delta t }^*  \circ \varphi_{ (\nicefrac{1}{2}- \alpha) \Delta t } \circ \varphi_{ \alpha \Delta t }^*,~
 y_2= (2 \sqrt{326} -36)^{\nicefrac{1}{3}}, ~ \alpha= \frac{y_2^2+6y_2-2}{12y_2}. 
\end{split}
\end{equation}
In the following four tests with varying initial conditions resulting in nonlinear Landau damping, 
the Weibel and the Weibel streaming instability with parameters according to \cite{kraus2016gempic,cheng2014discontinuous} are performed.
The second order splitting in eqn.~\eqref{spectral:symmsplit2ndlie} is used for the time discretization.
In most cases, the energy error is taken as a measure of correctness, yet the strength of the presented scheme is the preservation of structure, 
such that the energy error can be misleading, because the choice of a small enough time step, short simulation time and a sufficient resolution can
mimic conservation.
If the structure-preserving method is implemented correctly a simulation will exhibit long term stability, despite an insufficient resolution in time and space.
Here we also want to point out that the perfect energy conservation in~\cite{kraus2016gempic} for the Weibel instability was only achieved by high order integrators.
Stable results for low resolution are found in figs.~\ref{fig:spectral:vm1d2v:results_low},~\ref{fig:spectral:vm1d2v:results_low2},
and for better resolution in fig.~\ref{fig:spectral:vm1d2v:results_high},~\ref{fig:spectral:vm1d2v:results_high2},~\ref{fig:spectral:vm1d2v:results_high3}.
The default parameters are denoted in eqn.~\eqref{spectral:vm1d2v:params} along with the initial condition~\eqref{spectral:vm1d2v:initial}, which were adapted from~\cite{cheng2014discontinuous}.
\begin{figure}
\begin{equation}
\label{spectral:vm1d2v:params}
 \begin{matrix}
  \text{default} &   \epsilon,\beta_r,\beta_i,v_{0,1},v_{0,2},\delta,B_0=0,~c=1, \sigma_1,\sigma_2=1\\
  & ~N=N_x=N_{v_1}=N_{v_2}=32, ~       \Delta t = 0.05 \\
  & &\\
  \text{Strong Landau}  &     \epsilon_e=0.5,~ k=0.5,   \\
  &   v_{1,\max}=4.5 \sigma_1, ~v_{1,\min}=-v_{1,\max},~   v_{2,\max}=4.5 \sigma_2, ~v_{2,\min}=-v_{2,\max}\\
  \text{Weibel}  &      \beta_r=-10^{-3},~ k=1.25,~ \sigma_1=\frac{0.02}{\sqrt{2}},~  \sigma_2=\sqrt{12} \sigma_1,    \\
        &   v_{1,\max}=4.5 \sigma_1, ~v_{1,\min}=-v_{1,\max},~   v_{2,\max}=4.5 \sigma_2, ~v_{2,\min}=-v_{2,\max}\\
  \text{Weibel streaming sym.}& \sigma_1=\sigma_2=\frac{0.1}{\sqrt{2}},~k=0.2,~ \beta_i=10^{-3},~v_{0,1}=0.3,~v_{0,2}=-0.3,~ \delta=\frac{1}{2}\\
                               &   v_{1,\max}=0.9, ~v_{1,\min}=-v_{1,\max},~   v_{2,\max}=0.9, ~v_{2,\min}=-v_{2,\max}\\
  \text{Weibel streaming asym.}& \sigma_1=\sigma_2=\frac{0.1}{\sqrt{2}},~k=0.2,~ \beta_i=10^{-3},~v_{0,1}=0.5,~v_{0,2}=-0.1,~ \delta=\frac{1}{6}\\
                               &   v_{1,\max}=0.3~(or~0.7) , ~v_{1,\min}=-v_{1,\max},~   v_{2,\max}=1.05, ~v_{2,\min}=-0.55\\
 \end{matrix}
\end{equation}
\begin{equation}
\begin{split}
\label{spectral:vm1d2v:initial}
 f(x,v_1,v_2,t=0)&=  \frac{1+ \epsilon \cos(k x)}{2\pi \sigma_1 \sigma_2^2 }
  \expbb{ - \frac{v_1^2}{2\sigma_1^2}  }
 \left( 
 \delta \expbb{ - \frac{(v_2 -v_{0,1})^2}{2\sigma_2^2}  }
 +(1-\delta)\expbb{ - \frac{(v_2 -v_{0,2})^2}{2\sigma_2^2}}  
 \right)
 \\
 B_3(x,t=0) &= \beta_r \cos(k x ) + \beta_i \sin(k x)\\
 E_2(x,t=0)&= \alpha_r \cos(k x ) + \alpha_i \sin(k x)\\
 \partial_x E_1(x,t=0)&= 1 -  \int_{\mathbb{R}^2}  f(x,v_1,v_2,t) \dv
\end{split}
\end{equation}
\caption{Parameters and corresponding initial conditions for different Vlasov--Maxwell (1d2v) test-cases.
          The most challenging cases are the symmetric and asymmetric Weibel streaming instability.}
\end{figure}

\begin{figure}
 \centering
     \begin{subfigure}[t]{0.48\textwidth}
        \centering
     \includegraphics[width=\textwidth]{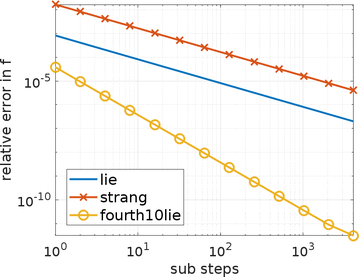}
     \caption{$L^2$ error}
    \end{subfigure}
    \enskip
    \begin{subfigure}[t]{0.48\textwidth}
        \centering
    \includegraphics[width=\textwidth]{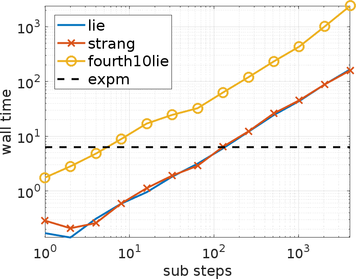}
    \caption{wall time}
    \end{subfigure}
\caption{By use of the matrix exponential $expm$ the split step $\mathcal{H}_{p_1}$ can be integrated exactly, but it is not matrix free and does at the moment not take advantage of the fast Fourier transform.
         But it can also be approximated by a sub stepped splitting, which is shown here for the asymmetric Weibel streaming instability at $t=t_{\max}=300$ in the fully nonlinear phase for $N_x=N_v=128$.
         Many sub-steps are required to approximate $\mathcal{H}_{p_1}$, such that high order methods are required (a) since the matrix exponential is comparably efficient (b).
         Nevertheless, experiments have shown that there was no visible difference in the fields for the presented test-cases when only two sub-steps where chosen.}
         \label{vm1d2vspectral:weibels:Hp1_approx}
\end{figure}


\begin{figure}
\begin{tabular}{c c c}
	\multicolumn{3}{c}{High resolution: $N_x=N_{v_1}=N_{v_3}=128$, $\Delta t=0.01$}\\
\hline
   \multicolumn{3}{c}{Weibel streaming instability (sym.)}\\
   \includegraphics[width=0.31\textwidth]{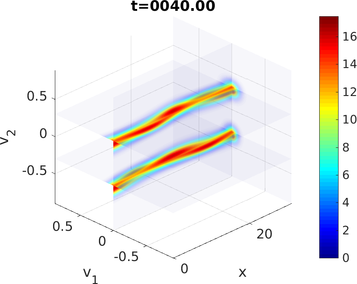}&
   \includegraphics[width=0.31\textwidth]{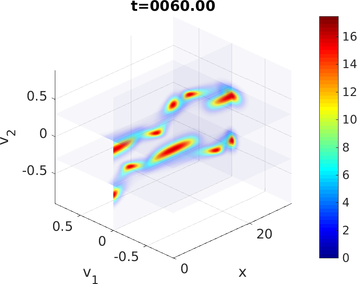}&
   \includegraphics[width=0.31\textwidth]{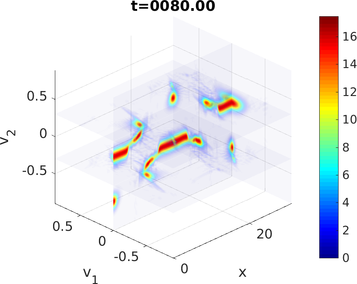}\\
   \includegraphics[width=0.31\textwidth]{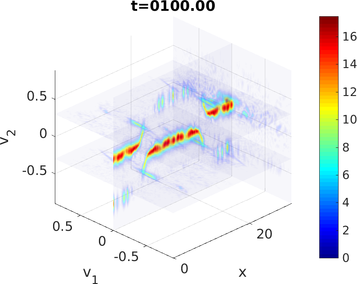}&
   \includegraphics[width=0.31\textwidth]{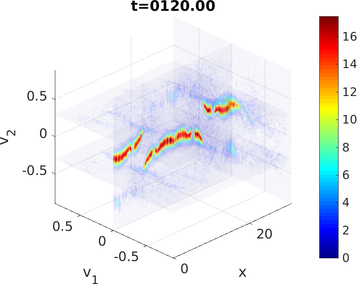}&
   \includegraphics[width=0.31\textwidth]{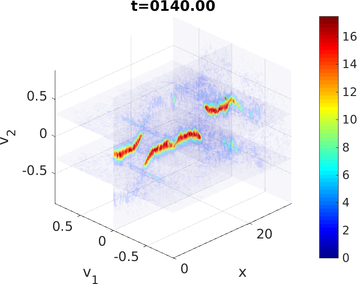}\\
  \hline 
\multicolumn{3}{c}{Weibel streaming instability (asym.)}\\
   \includegraphics[width=0.31\textwidth]{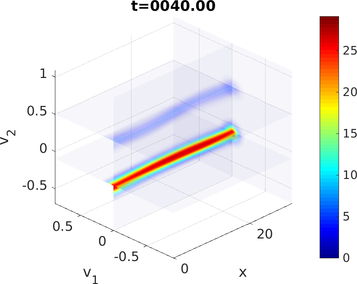}&
   \includegraphics[width=0.31\textwidth]{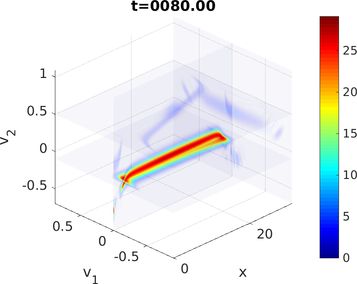}&
   \includegraphics[width=0.31\textwidth]{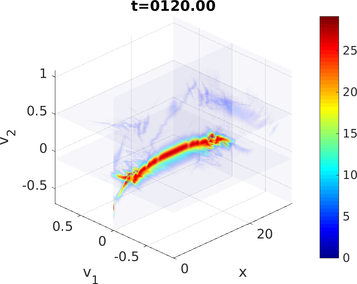}\\
   \includegraphics[width=0.31\textwidth]{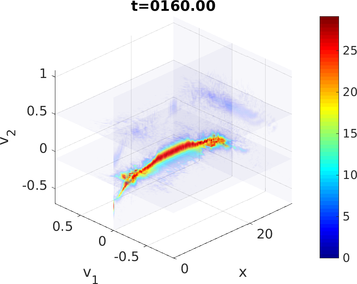}&
   \includegraphics[width=0.31\textwidth]{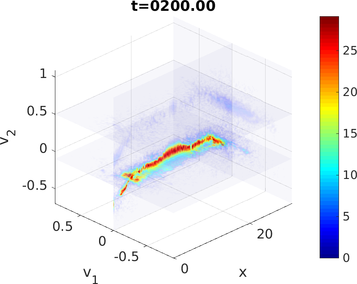}&
   \includegraphics[width=0.31\textwidth]{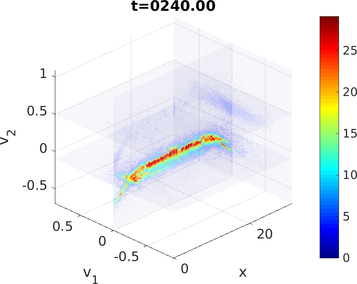}\\
\end{tabular}
\caption{Phase space densities for Vlasov--Maxwell 1d2v simulations under high resolution.}
\end{figure}

\FloatBarrier
\section{Summary}
This article introduced new Fourier splitting methods for the four-dimensional magnetized Vlasov--Poisson system and the three dimensional Vlasov--Maxwell system with new splitting schemes under challenging test cases. For a very strong magnetic field, the fluid model is an appropriate approximation to the kinetic model and contains all the relevant dynamics, hence it makes less sense to use the kinetic model in this case, as the fluid model
will always be much cheaper. The most prominent examples are asymptotic preserving schemes\cite{filbet2016asymptotically,filbet2016asymptotically} which are mainly designed to preserve the asymptotic model
but at the expense of discretising the entire kinetic phase space. The best scheme presented here (Scovel's method) performs independently of the strength of the magnetic field very well.
But it has to be pointed out that its intended use is for an in-between scenario, where it is unclear whether the asymptotic model is already suitable.\\
In a next step the exponential Boris and Scovel splitting can be adapted for the Vlasov--Amp\`{e}re and Vlasov--Maxwell equations. There the exponential Boris, which does not require a homogeneous magnetic field
is merely a slightly different splitting and all the necessary formulas are already presented here. 
For Scovel's method, the critical point is the exact integration of the Amp\`{e}re equation over time. Eventually, this can be solved in general using Bessel functions for the gyroaverage over one period of rotation,
similar to the Bessel functions appearing in gyrokinetic theory~\cite{steiner2015gyroaverage}, and Gauss' quadrature for the remainder.
In the future, an interesting test case for a four-dimensional Vlasov--Maxwell system under a strong field could be kinetic shear Alfven waves~\cite{dannert2004vlasov}, which contains a nonhomogenous magnetic field. 
In Fourier space, we are limited to constant-coefficient advection, such that all the splitting schemes presented here have the one-dimensional advection as the underlying building block. In principle the Semi-Lagrangian 
method does not have such a limitation if a full-dimensional interpolation is used. But for highly scalable codes as e.g.~\cite{kormann2019massively}
this is not the case, because for performance reasons only one dimensional interpolation is implemented.
Therefore, those codes can also benefit from the novel schemes presented here.

\section{Acknowledgement}
This work has been carried out within the framework of the EUROfusion Consortium and has received funding from the Euratom research and training programme 2014-2018 and 2019-2020 under grant agreement No 633053. The views and opinions expressed herein do not necessarily reflect those of the European Commission.

 \printbibliography
\end{document}